# Title page

Title: Simultaneous optimization of non-coplanar beam orientations and cumulative EQD2 distribution for high-dose reirradiation of locoregionally recurrent non-small cell lung cancer

Authors and affiliations: Nathan Torelli[1,*], Jonas Willmann[1,2], Katja Dähler[1], Madalyne Day[1], Nicolaus Andratschke[1], Jan Unkelbach[1]

[1] Department of Radiation Oncology, University Hospital Zurich and University of Zurich, Zurich, Switzerland

[2] Department of Medical Physics, Memorial Sloan Kettering Cancer Center, New York, NY, USA

* Corresponding author, email: nathan.torelli@usz.ch

# Abstract

Background and Purpose: Reirradiation for non-small cell lung cancer (NSCLC) is commonly delivered using coplanar techniques. In this study, we developed a beam orientation optimization algorithm for reirradiation planning to investigate whether the selection of favorable non-coplanar beam orientations may limit cumulative doses to critical organs-at-risk (OARs) and thus improve the therapeutic window.

Materials and Methods: Fifteen cases of challenging high-dose reirradiation for locoregionally recurrent NSCLC were included in this in-silico study. For each patient, the dose distribution from the previous treatment was first mapped to the reirradiation planning CT using rigid dose registration, and subsequently converted to equivalent dose in 2 Gy fractions (EQD2). A 2-arc non-coplanar reirradiation plan, combining dynamic gantry *and* couch rotation, was then generated using an EQD2-based direct aperture optimization algorithm, which allows for the simultaneous optimization of the dynamic gantry-couch path and the cumulative EQD2 distribution. Non-coplanar reirradiation plans were benchmarked against 2-arc coplanar VMAT plans, which mimic state-of-the-art practice for reirradiation of NSCLC.

Results: Non-coplanar reirradiation plans could reduce the maximum cumulative EQD2 to critical OARs such as bronchial tree, esophagus, thoracic wall and trachea by at least 5 $Gy_2$ for 6 out of 15 patients compared to coplanar reirradiation plans. At the same time, target coverage and lung EQD2 metrics were comparable for both methods.

Conclusions: The automated selection of favorable non-coplanar beam orientations may reduce the maximum cumulative EQD2 to critical OARs in challenging thoracic reirradiation cases. This allows to explore either better OAR sparing or dose-escalation in future clinical studies.

# 1. Introduction

Non-small cell lung cancer (NSCLC) is the most common lung cancer subtype occurring in 85% of patients and one of the leading causes of cancer death worldwide [1,2]. Despite technological advances and improvements in treatment paradigms, about 16-37% of NSCLC patients will develop locoregional recurrence, with 2- and 5-year survival of 21% and 6%, respectively [3-6]. Reirradiation is the main curative-intent treatment option available for these patients [7]. Retrospective studies suggest that local control and even prolonged survival can be achieved with reirradiation in selected patients [6-10], particularly when using highly conformal treatment techniques and dose fractionation strategies that balance efficacy with safety. However, reirradiation also poses a significant therapeutic challenge due to high cumulative doses which may lead to severe treatment-related toxicities [11].

In this context, it is very important to create reirradiation plans that balance tumor control and the risk of toxicity from cumulative doses. Yet, no commercial treatment planning system currently exists that allows for the optimization of reirradiation plans based on the dose delivered in the previous treatment, while also accounting for normal tissue recovery and dose fractionation. In current clinical practice, reirradiation planning is still mostly based on a manual trial-and-error approach, in which a reirradiation plan is first generated *independently* of the previous dose distribution, and the cumulative equivalent dose in 2 Gy fractions (EQD2) distribution from all treatments combined is only evaluated a posteriori [12]. The reirradiation plan may then be iteratively adjusted if some clinical goals on the cumulative EQD2 are not met. This approach is not only very time consuming, but may also lead to suboptimal reirradiation plans. Therefore, approaches to facilitate and improve reirradiation planning are warranted [13].

Recently, few groups investigated approaches to improve reirradiation planning by enabling EQD2-based treatment plan optimization. Murray *et al*, for example, developed a reirradiation planning workflow which allows to generate reirradiation plans based on the voxel-by-voxel cumulative EQD2 of the previous treatment and the reirradiation plan [14]. They demonstrated that using such an approach, clinically acceptable reirradiation plans could be generated which required less constraint relaxation or allowed for dose escalation. Similarly, Meyer *et al* developed a reirradiation planning solution that leverages a combination of dose accumulation scripts to define residual dose constraints for the critical OARs based on cumulative EQD2 isodose line structures, and an automated optimization algorithm [15]. Using such a workflow, they could generate reirradiation plans with superior quality compared to manually generated plans.

In this study, we aimed to expand on reirradiation planning research by investigating whether reirradiation treatments can be improved by selecting favorable non-coplanar beam orientations that limit dose accumulation to critical OARs. In most situations, in fact, coplanar radiotherapy techniques are often used for both the initial and the reirradiation treatments [16]. However, when a tumor recurs very close or even overlaps with the previously irradiated target volume, the use of coplanar beam orientations will unavoidably accumulate dose to already exposed body regions. To address this problem, we developed a method which allows to simultaneously optimize non-coplanar dynamic trajectories to be used in the reirradiation plan and the cumulative EQD2 distribution of both the initial and the reirradiation plans. Non-coplanar dynamic trajectories consist of a technique that combines dynamic gantry and couch rotation, and represent an efficient approach to deliver non-coplanar radiotherapy treatments using conventional C-arm linear accelerators [17-25]. The proposed method has been evaluated for several challenging high-dose reirradiation cases with locoregional recurrent NSCLC.

## 2. Materials and Methods

### 2.1 Patients

Fifteen cases of challenging high-dose reirradiation for locoregionally recurrent NSCLC, whose treatment details are reported in Table 1, were retrospectively considered in this study. All patients were treated for tumor recurrence at our institution between 2016 and 2024 using a coplanar volumetric modulated arc therapy (VMAT) technique.

### 2.2 Reirradiation planning workflow

Reirradiation plans were generated using an in-house developed treatment planning workflow, which is schematically illustrated in Figure 1 and further detailed in the following.

First, the dose distribution from the initial treatment was mapped to the reirradiation planning CT using proprietary dose registration algorithms of the commercial treatment planning system Eclipse (Varian, A Siemens Healthineers Company). In this study, rigid dose registration was used to mimic the approach adopted in our clinic for generating the reirradiation plans for the considered NSCLC patients. However, the same planning workflow may also be used in combination with deformable dose registration. Subsequently, the registered physical dose distribution was converted to EQD2 distribution as follows:

$$e_i^{(pre)} = \frac{d_i^{(pre)} \left[ (\alpha/\beta)_i + d_i^{(pre)} \right]}{2 \left[ (\alpha/\beta)_i + 2 \right]} \qquad \forall i \qquad (1)$$

where $e_i^{(pre)}$ and $d_i^{(pre)}$ are the EQD2 and physical dose in voxel $i$ from the initial treatment(s), respectively, and $(\alpha/\beta)_i$ is the α/β-ratio of the tissue that voxel $i$ belongs to. Following our institutional guidelines, $\alpha/\beta = 2\ Gy$ was set for the spinal cord and $\alpha/\beta = 3\ Gy$ was set for the rest of the body. A non-coplanar reirradiation plan, combining dynamic gantry *and* couch rotation, was finally generated using a novel EQD2-based direct aperture optimization (DAO) algorithm, which allowed for the simultaneous optimization of the dynamic gantry-couch path and the cumulative EQD2 distribution of both the initial and the reirradiation plans.

## 2.3 EQD2-based direct aperture optimization algorithm

The proposed DAO algorithm, which was first suggested by Torelli and Unkelbach [26] and extended to the reirradiation setting in this study, combined a column generation based method to iteratively add apertures from promising non-coplanar beam orientations and a gradient-based method to refine the weights and shapes of all apertures along a non-coplanar dynamic arc. Starting from an empty set of control points, the most promising aperture shapes at each candidate beam orientation $b \in B$ were determined by minimizing the sum of partial derivatives of an objective function $f$ (evaluated for the cumulative EQD2) with respect to the fluence of all bixels contained in the aperture:

$$\underset{x_L, x_R}{\text{minimize}} \sum_{l \in L_b} \sum_{p=x_L^l}^{x_R^l} \frac{\partial f}{\partial v^{lp}} \qquad (2)$$

Here, $v^{lp}$ refers to the fluence of the bixel in leaf pair $l$ at position $p \in \{x_L^l, \ldots, x_R^l\}$, $x_L^l$ and $x_R^l$ are the left and right positions of the leaf pair $l$, respectively, and $L_b$ is the set of all MLC leaf pairs that apertures in field $b$ can utilize. Descriptions of strategies to solve the so-called pricing problem in Equation (2) can be found in the work of Romeijn *et al* [27] and Men *et al* [28]. The resulting candidate apertures had an associated price, given by the sum of partial derivatives of the objective function $f$ with respect to all bixels not covered by the multileaf collimator (MLC). Since a negative partial derivative indicated that increasing the fluence of the bixel decreased the objective function value, the aperture delivered from the beam orientation with the lowest price was added to the treatment plan at each iteration. In this study, the set of candidate beam orientations $B$ encompassed beam orientations at different pairs of gantry and couch angles, with the gantry angles ranging from -180° to +180° in 5°-steps and the couch angles ranging from -90° to +90° in 2.5°-steps. The treatment isocenter was selected as the center-of-mass of the planning target volume. Beam orientations that may lead to a collision between the gantry and the system composed by couch and patient have been determined for general patient anatomies and treatment isocenters (i.e. they were not patient-specific), and were excluded from the set of candidate beam orientations. After the addition of each new aperture, the search space for candidate beam orientations was updated such that only apertures could be added at beam orientations that could still be efficiently reached by the dynamic gantry-couch path, without the need to slow down the gantry rotation speed. Therefore, the non-coplanar arcs in this study featured similar delivery times as coplanar arcs.

After the addition of each new aperture to the reirradiation plan, the weights of all the already added apertures were optimized and the apertures' shapes were refined. To this end, the following optimization problem was solved:

$$\underset{(x_L, x_R), \omega}{\text{minimize}} \quad f(\tilde{e}) \tag{3}$$

$$\text{subject to} \quad \tilde{e}_i = \frac{d_i[(\alpha/\beta)_i + d_i]}{2[(\alpha/\beta)_i + 2]} + e_i^{(pre)} \quad \forall i \tag{4}$$

$$d_i = \sum_{k \in K} \omega_k \sum_{l \in L_k} \Phi_i^{kl}(x_L^{kl}, x_R^{kl}) \quad \forall i, \forall t \tag{5}$$

$$\omega^k \geq 0 \quad \forall k \tag{6}$$

$$\left| x_L^{(k-1)l} - x_L^{kl} \right| \leq \Delta x_{max} \quad \forall k \geq 1, \forall l \tag{7}$$

$$\left| x_R^{(k-1)l} - x_R^{kl} \right| \leq \Delta x_{max} \quad \forall k \geq 1, \forall l \tag{8}$$

$$x_L^{kl} \leq x_R^{kl} \quad \forall k, \forall l \tag{9}$$

where $\tilde{e}_i$ is the cumulative EQD2 to voxel $i$, $d_i$ is the physical dose delivered to voxel $i$ by all MLC-based apertures $k \in K$ in the reirradiation plan, $\Phi_i^{kl}$ is the dose contribution of the $l$-th leaf pair of aperture $k$ to voxel $i$ per unit intensity and $\omega_k$ is the MU weight of aperture $k$. The parameters $x_L^{kl}$ and $x_R^{kl}$ describe the positions of the left and right MLC leaves in the $l$-th leaf pair of aperture $k$, respectively, and $L_k$ is the set of all MLC leaf pairs in aperture $k$. The constraints in Equations (7)-(8) were used to limit the MLC displacement between neighboring control points by a maximum distance:

$$\Delta x_{max} = \max\left( v_{max} \frac{\Delta\theta}{\omega_g}, v_{max} \frac{\Delta\phi}{\omega_c} \right) \tag{10}$$

where $\Delta\theta$ and $\Delta\phi$ denote the difference in gantry and couch angles between neighboring control points, $v_{max}$ is the maximum MLC leaf speed and $\omega_g$ and $\omega_c$ are the maximum rotation speed of the gantry and the couch, respectively. In this study, these parameters have been set to $v_{max} = 2.5$ cm/s, $\omega_g = 6°/s$ and $\omega_c = 3°/s$, respectively, which correspond to the dynamic parameters of a TrueBeam treatment unit equipped with a Millennium MLC 120 (Varian, A Siemens Healtineers Company). This guaranteed an efficient treatment delivery, as described by Peng *et al* [29]. The optimization problem in Equations (3)-(9) was solved using a gradient-based DAO approach inspired by the work of Cassioli and Unkelbach [30] in combination with our in-house implementation of the L-BFGS quasi-Newton algorithm [31].

## 2.4 In-silico evaluation of non-coplanar reirradiation plans

For each patient, a 2-arc non-coplanar reirradiation plan was generated using the proposed planning workflow. This plan was then benchmarked against a 2-arc coplanar reirradiation plan, which mimicked current state-of-the-art practice for reirradiation of NSCLC patients and reflected the technique that was used to treat the considered patients in the clinics. Because of differences in both the dose calculation and plan optimization algorithms used in the commercial and research treatment planning systems, a direct comparison of the non-coplanar reirradiation plan with the clinically delivered reirradiation plan was not meaningfully possible. Therefore, the clinically delivered coplanar reirradiation plan was mimicked in our research treatment planning system by using the same EQD2-based DAO algorithm as for the non-coplanar reirradiation plan, but fixing the beam orientations to $B_0 = \{(\theta_k, \phi_k) | \theta_k = -180° + k * 5°, \phi_k = 0° \: \forall k \in \{0, 1, ..., 72\}\}$.

For the purpose of this study, both the coplanar and non-coplanar reirradiation plans were forced to achieve a similar target coverage, in order to quantify the difference in between the two approaches in terms of OAR dose reduction. For each patient, the objective function in Equation (3) was defined through a weighted sum of dose-volume, mean dose and dose conformity objectives (as reported in the Supplementary material A). To guarantee that the prescribed dose was delivered to the tumor in the reirradiation plan and that the reirradiation plan was conformal, the under- and over-dose planning objectives for the PTV, as well as the normal tissue objective, were evaluated only for the EQD2 distribution resulting from the reirradiation plan, and not for the cumulative EQD2. The comparison between the coplanar and non-colpanar reirradiation plans was performed by evaluating target coverage, dose conformity and cumulative EQD2 metrics to the OARs (without considering any overlap with the PTV).

## 3 Results

The greatest reduction of OAR doses between the coplanar and non-coplanar reirradiation plans was observed for patient 4, who had a tumor recurrence very close (but without overlap) to a few OARs that already received a large dose in the initial treatment (Figure 2). In this situation, the use of coplanar VMAT arcs unavoidably deposited additional radiation dose to OAR regions that already received a very large dose in the initial radiotherapy. Consequently, the coplanar reirradiation plan achieved very high maximum EQD2 of 141.3 $Gy_2$ in the bronchial tree, 95.9 $Gy_2$ in the esophagus, 120.6 $Gy_2$ in the great vessel and 127.6 $Gy_2$ in the trachea, which exceeded the maximum tolerated EQD2 constraints (Table 2). By using optimized non-coplanar beam orientations, on the other hand, the non-coplanar reirradiation plan could better direct the integral dose to regions of the body which previously did not receive any dose or only limited dose from the initial radiotherapy, while also achieving a much steeper dose gradient outside of the target. Compared to the coplanar reirradiation plan, the non-coplanar reirradiation plan reduced the maximum cumulative EQD2 to bronchial tree, esophagus, great vessel and trachea by 9.0 $Gy_2$ (-6.4%), 5.0 $Gy_2$ (-5.2%), 5.6 $Gy_2$ (-4.6%) and 4.9 $Gy_2$ (-3.8%), respectively, for a similar target coverage (CI: 0.75 vs 0.75).

In 6 out of 15 patients (40%), the non-coplanar reirradiation plan could reduce the maximum cumulative EQD2 to at least one critical OAR by at least 5 $Gy_2$ compared to the coplanar reirradiation plan (Figure 3). In particular, large reductions in the maximum cumulative EQD2 were achieved for bronchial tree (up to -9.0 $Gy_2$), esophagus (up to -5.8 $Gy_2$), heart (up to -6.4 $Gy_2$), thoracic wall (up to -7.5 $Gy_2$), trachea (up to -5.3 $Gy_2$), great vessel (up to -5.6 $Gy_2$) and brachial plexus (up to -9.4 $Gy_2$). At the same time, the mean lung EQD2 was reduced on average by 0.2±0.1 $Gy_2$ ([-0.5 $Gy_2$,-0.1 $Gy_2$]), lung V5$Gy_2$ was reduced on average by 0.6±1.0**%** ([-3.5%,+0.3%]) and lung V20$Gy_2$ was reduced on average by 0.7±0.7% ([-2.7%,-0.1%]) using optimized non-coplanar versus coplanar arcs (more detailed dosimetric results for each individual patientare reported in the Supplementary material B).

Reduction of the cumulative EQD2 to critical OARs required less constraint relaxation. For 8 out of 15 of the considered patients (53%), in fact, the coplanar reirradiation plans exceeded the cumulative EQD2 constraints for at least one critical OAR (Table 2). While in most of the cases the OAR constraints were already exceeded by the initial radiotherapy alone (mainly due to overlaps between the previous target volume and critical OARs), in 6 out of 18 OAR constraint violations (33%) this was due to the additional dose contribution from the coplanar reirradiation plan. By using optimized non-coplanar arcs, the non-coplanar reirradiation plans could better spare some of these OARs. In particular, the maximum EQD2 to the great vessel in patient 4 could be kept below the recommended EQD2 constraint of 120 Gy$_2$ with the non-coplanar reirradiation plan. Similarly, the maximum EQD2 to the bronchial tree in patient 1, to the esophagus in patient 4 and to the trachea in patient 10 could be kept within +2% from the corresponding constraints.

## 4 Discussion

In this study, an EQD2-based beam orientation optimization algorithm for reirradiation planning was developed and demonstrated for several cases of challenging high-dose reirradiation for locoregionally recurrent NSCLC. By selecting appropriate non-coplanar dynamic trajectories to be delivered in the reirradiation plan, it was shown that the integral dose could be directed to regions of the body that did not receive any dose or only a little dose in the initial radiotherapy, thus considerably reducing the maximum cumulative EQD2 to critical thoracic OARs compared to reirradiation plans that were delivered using coplanar VMAT arcs. This may limit treatment-related toxicities or allow for tumor dose escalation.

This study examined two concepts which are currently not released in commercial treatment planning systems: plan optimization based on cumulative EQD2 constraints and treatment delivery using non-coplanar dynamic trajectories. Murray *et al* [14] and Meyer *et al* [15] already showed that EQD2-based treatment plan optimization is an efficient and effective approach for generating reirradiation plans, which can improve on manual trial-and-error planning methods. However, while a lot of research is currently focused on making EQD2-based reirradiation planning available on commercial treatment planning systems [13], in this study it was demonstrated that EQD2-based treatment plan optimization alone may not always suffice to create clinically acceptable reirradiation plans, especially for challenging high-dose reirradiation patients. Approaches to improve reirradiation plans by further limiting the cumulative OAR exposure, such as the use of non-coplanar beam orientations presented in this study or the use of proton therapy [32,33], are indeed equally important to facilitate reirradiation by potentially reducing the risk of treatment-related toxicities and allowing for dose escalation to the target. The delivery of non-coplanar dynamic trajectories was also previously investigated by several authors [17-26]. Previous studies demonstrated for several treatment sites that non-coplanar arcs combining dynamic gantry and couch rotation could considerably reduce the dose to OARs, while maintaining a similar delivery efficiency as coplanar VMAT arcs. In most studies, however, the selection of favorable non-coplanar beam orientations was mainly dictated by geometrical reasons (i.e. to avoid delivering dose directly through critical OARs). Per contra, the beam selection strategy used in this study aimed to spare regions of the body which already received a large dose in the initial treatment and was thus less influenced by the patient geometry (similar to the approach adopted by Bedford et al [34], who considered lung function).

The reirradiation planning framework proposed in this work was implemented within our in-house research treatment planning system and thus no direct clinical implementation is possible yet. In addition, although the simultaneous rotation of gantry and couch is supported by most commercial C-arm linear accelerators, the clinical implementation of non-coplanar dynamic arcs still faces some potential problems, like the increased risk of collision between the gantry head and the system composed by couch and patient, and intra-fraction patient motion due to the continuous couch rotation. In that regard, the development of more accurate collision prediction models and the use of adequate immobilization devices may be warranted for a clinical use of non-coplanar dynamic trajectories. Nevertheless, alternative approaches for the generation and delivery of non-coplanar reirradiation plans which present less hurdles for a clinical implementation may be considered. For example, dose prediction algorithms could be used in combination with dose accumulation scripts to determine fixed couch kicks for which a non-coplanar VMAT reirradiation plan is expected to minimize the cumulative EQD2 exposure of the most critical OARs. Alternatively, coplanar VMAT arcs may also be combined with few non-coplanar beams as previously shown by Sharfo *et al* [35,36].

Further research must be conducted to identify which patients may benefit the most from the use of optimized non-coplanar beam orientations, e.g. depending on the size and location of the tumor with respect to the most critical OARs in the initial and reirradiation treatments. In addition, EQD2-based beam orientation optimization may be investigated also for other treatment sites or for reirradiation using proton therapy.

# Tables

*Table 1: Details of the initial and reirradiation treatments for all considered patients. Tumor location is described as ultra-central (UC, when the PTV overlaps with bronchial tree, trachea or esophagus), central (C, when the PTV is at a distance lower than 2 cm from bronchial tree, trachea or esophagus) or peripheral (P, when the PTV is at a distance larger than 2 cm from bronchial tree, trachea or esophagus). Type I reirradiation refers to cases in which there is geometrical overlap between the target volumes in the reirradiation and previous treatments, whereas type II reirradiation describes situations in which there are concerns about cumulative dose toxicity, but no geometrical overlap of the PTV volumes. For patients 1, 3-4, 6, 8-11 and 13, the reirradiation under consideration is the third course of radiotherapy.*

| | Previous treatment(s) | | | Reirradiation treatment | | | | | |
|---|---|---|---|---|---|---|---|---|---|
| | Tumor location | PTV volume (cm³) | Fractionation scheme | Time since previous treatment (months) | Reirradiation type | Tumor location | PTV volume (cm³) | Fractionation scheme | Dose-limiting OARs |
| **Patient 1** | UC | 288 | 33 x 2 Gy | 15 | I | UC | 49 | 10 x 2.75 Gy | Bronchial tree |
| | UC | 17 | 10 x 4.5 Gy | 4 | | | | | |
| **Patient 2** | P | 98 | 3 x 12.5 Gy | 16 | I | UC | 238 | 30 x 2.2 Gy | Esophagus |
| **Patient 3** | UC | 179 | 30 x 2.2 Gy | 32 | I | UC | 10 | 5 x 5 Gy | Esophagus |
| | P | 17 | 10 x 4.5 Gy | 6 | | | | | |
| **Patient 4** | UC | 432 | 35 x 2 Gy | 19 | I | C | 15 | 10 x 4.5 Gy | Bronchial tree, esophagus, great vessel, trachea |
| | UC | 66 | 30 x 2 Gy | 17 | | | | | |
| **Patient 5** | UC | 603 | 30 x 2 Gy | 4 | II | C | 12 | 5 x 7 Gy | Lungs |
| **Patient 6** | UC | 90 | 33 x 2.2 Gy | 41 | I | UC | 489 | 5 x 4 Gy | Bronchial tree, esophagus, heart, great vessel |
| | UC | 76 | 22 x 2.75 Gy | 29 | | | | | |
| **Patient 7** | C | 226 | 30 x 2.2 Gy | 47 | II | P | 20 | 5 x 10 Gy | Thoracic wall |
| **Patient 8** | C | 14 | 20 x 3 Gy | 14 | I | P | 59 | 8 x 5 Gy | Thoracic wall |
| | P | 11 | 5 x 9 Gy | 13 | | | | | |
| **Patient 9** | UC | 525 | 31 x 2 Gy | 61 | I | UC | 288 | 30 x 2 Gy | Brachial plexus, trachea |
| | C | 17 | 8 x 6 Gy | 42 | | | | | |
| **Patient 10** | UC | 525 | 31 x 2 Gy | 8 | I | UC | 12 | 30 x 2 Gy | Bronchial tree, trachea |
| | P | 17 | 5 x 7 Gy | 8 | | | | | |
| **Patient 11** | UC | 149 | 33 x 2 Gy | 6 | II | C | 175 | 20 x 2.75 Gy | Trachea |
| | P | 14 | 5 x 10 Gy | 6 | | | | | |
| **Patient 12** | UC | 1015 | 27 x 2 Gy | 12 | I | UC | 48 | 10 x 4.85 Gy | Bronchial tree |
| **Patient 13** | UC | 372 | 27 x 2.2 Gy | 49 | I | UC | 307 | 10 x 4.5 Gy | Bronchial tree |
| | C | 113 | 20 x 2.75 Gy | 12 | | | | | |
| **Patient 14** | UC | 175 | 19 x 2.75 Gy | 95 | II | C | 155 | 22 x 2.75 Gy | Heart |
| **Patient 15** | UC | 904 | 30 x 2 Gy | 18 | I | UC | 232 | 18 x 2.75 Gy | Bronchial tree, heart |

*Table 2: Cumulative EQD2 metrics that exceeded the recommended EQD2 constraints (according to our institutional clinical protocol for thoracic reirradiation), both prior to reirradiation and after reirradiation using the coplanar and non-coplanar plans. The cumulative EQD2 metrics for all patients and OARs (including the ones respecting the constraints) are reported in the Supplementary material B. Percentage differences in EQD2 values between the coplanar and non-coplanar reirradiation plans are reported for each parameter.*

|  | OAR | Recommended EQD2 constraint | Previous treatment(s) (Gy$_2$) | Coplanar reirradiation plan (Gy$_2$) | Non-coplanar reirradiation plan (Gy$_2$) |
|---|---|---|---|---|---|
| **Patient 1** | Bronchial tree | $D_{max}$ < 120 Gy$_2$ | 71.4 | 124.0 | 122.5 (-1.2%) |
|  | Esophagus | $D_{max}$ < 90 Gy$_2$ | 126.9 | 126.9 | 126.9 (=) |
|  | Trachea | $D_{max}$ < 110 Gy$_2$ | 117.2 | 117.2 | 117.2 (=) |
| **Patient 3** | Esophagus | $D_{max}$ < 90 Gy$_2$ | 98.8 | 100.6 | 99.7 (-0.9%) |
| **Patient 4** | Bronchial tree | $D_{max}$ < 120 Gy$_2$ | 120 | 141.3 | 132.3 (-6.4%) |
|  | Esophagus | $D_{max}$ < 90 Gy$_2$ | 86.3 | 95.9 | 90.9 (-5.2%) |
|  | Great vessel | $D_{max}$ < 120 Gy$_2$ | 114.9 | 120.6 | 115.0 (-4.6%) |
|  | Trachea | $D_{max}$ < 110 Gy$_2$ | 117.5 | 127.6 | 122.7 (-3.8%) |
| **Patient 6** | Bronchial tree | $D_{max}$ < 120 Gy$_2$ | 131.3 | 142.4 | 142.7 (=) |
|  | Esophagus | $D_{max}$ < 90 Gy$_2$ | 92.4 | 99.5 | 95.8 (-3.7%) |
|  | Heart | $D_{max}$ < 85 Gy$_2$ | 144.7 | 159.2 | 157.4 (-1.1%) |
|  | Great vessel | $D_{max}$ < 120 Gy$_2$ | 143.5 | 159.6 | 158.5 (-0.7%) |
| **Patient 8** | Thoracic wall | $D_{max}$ < 120 Gy$_2$ | 237.8 | 270.9 | 270.3 (-0.2%) |
| **Patient 10** | Bronchial tree | $D_{max}$ < 120 Gy$_2$ | 70.4 | 121.2 | 121.2 (=) |
|  | Thoracic wall | $D_{max}$ < 120 Gy$_2$ | 173.9 | 173.9 | 173.9 (=) |
|  | Trachea | $D_{max}$ < 110 Gy$_2$ | 71.3 | 115.1 | 111.4 (-3.2%) |
| **Patient 13** | Bronchial tree | $D_{max}$ < 120 Gy$_2$ | 127.4 | 129.2 | 129.0 (=) |
| **Patient 15** | Heart | $D_{max}$ < 85 Gy$_2$ | 60.6 | 89.4 | 89.1 (=) |

Figures

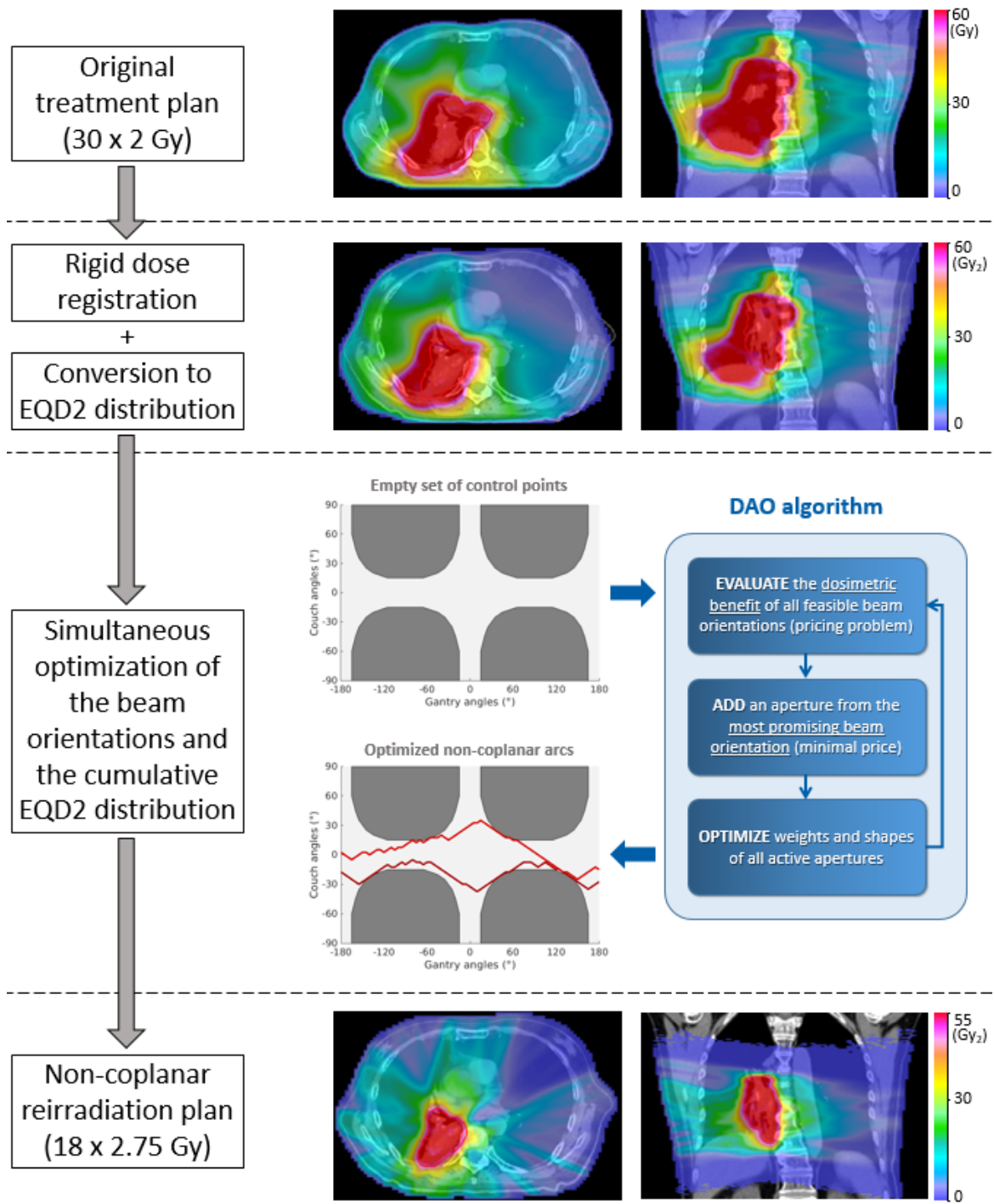

Figure 1: Schematic illustration of the reirradiation planning workflow used in this study. After the dose distribution of the initial radiotherapy has been registered to the reirradiation planning CT and subsequently converted to EQD2, a direct aperture optimization algorithm is used to simultaneously optimize the beam orientations for the reirradiation plan and the cumulative EQD2 distribution. The reirradiation plan delivers non-coplanar arcs, which combine dynamic gantry and couch rotation, aiming to minimize dose accumulation to critical OARs. The dark grey regions in the gantry-couch map represent beam orientations which may lead to collisions between the gantry and the system composed by couch and patient, and are therefore excluded from the search space.

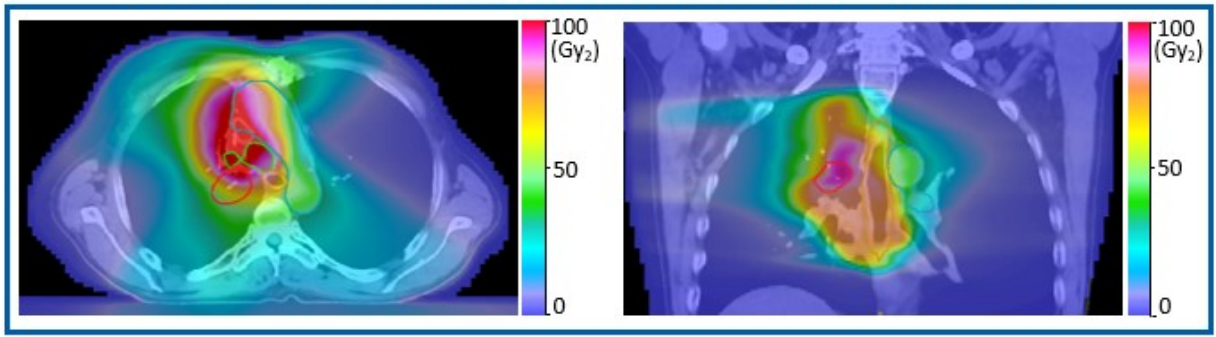
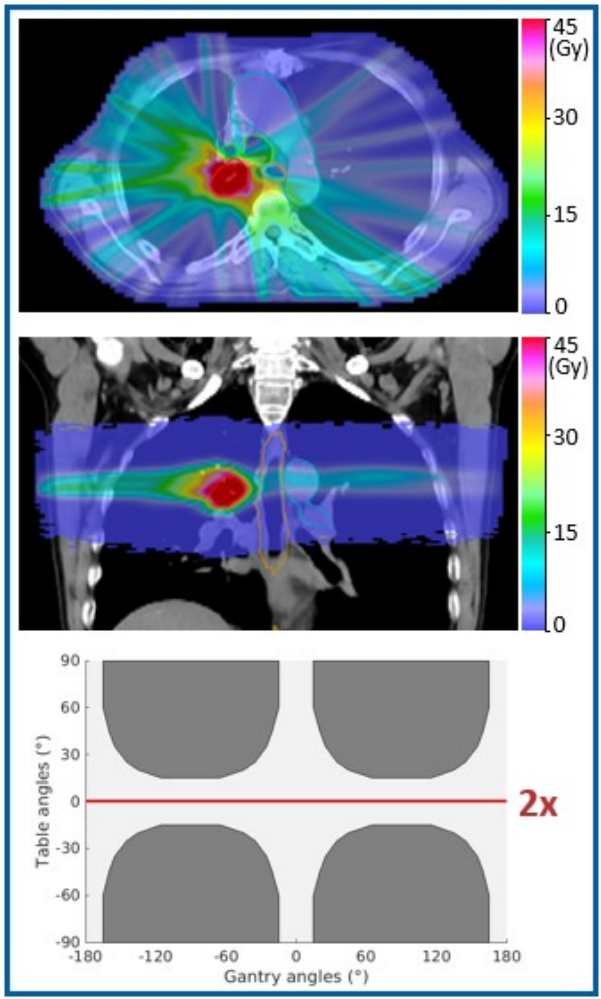
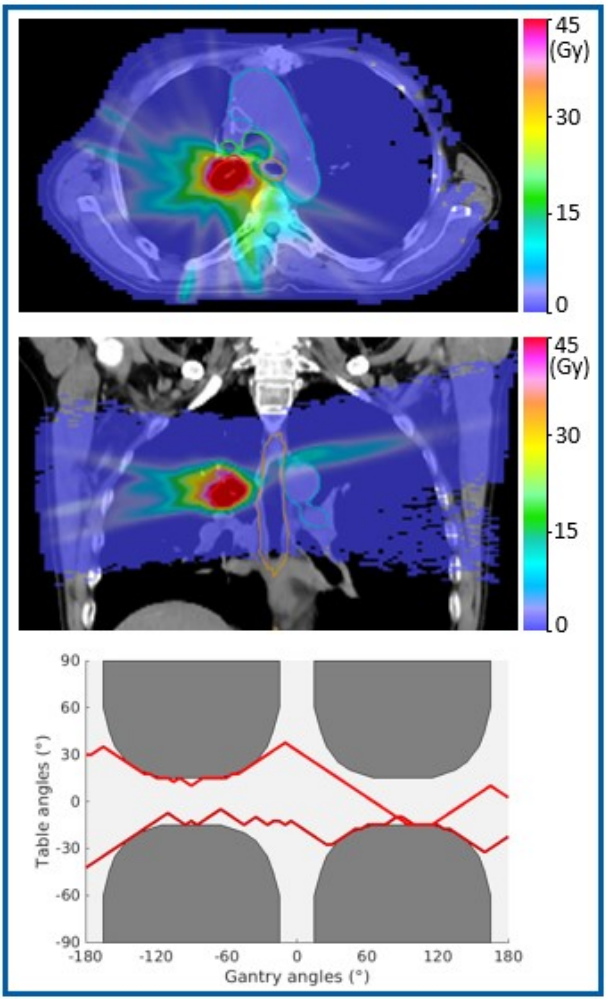

*Figure 2: Comparison of the coplanar and non-coplanar reirradiation plans generated for patient 4. Contours of PTV (red), bronchial tree (green), esophagus (brown) and great vessel (blue) are delineated in both the transversal and coronal planes of the reirradiation scan. The gantry-couch paths for both reirradiation plans are also shown, where dark grey regions indicate beam orientations leading to collision between gantry and couch (and are therefore excluded from the set of candidate beam orientations).*

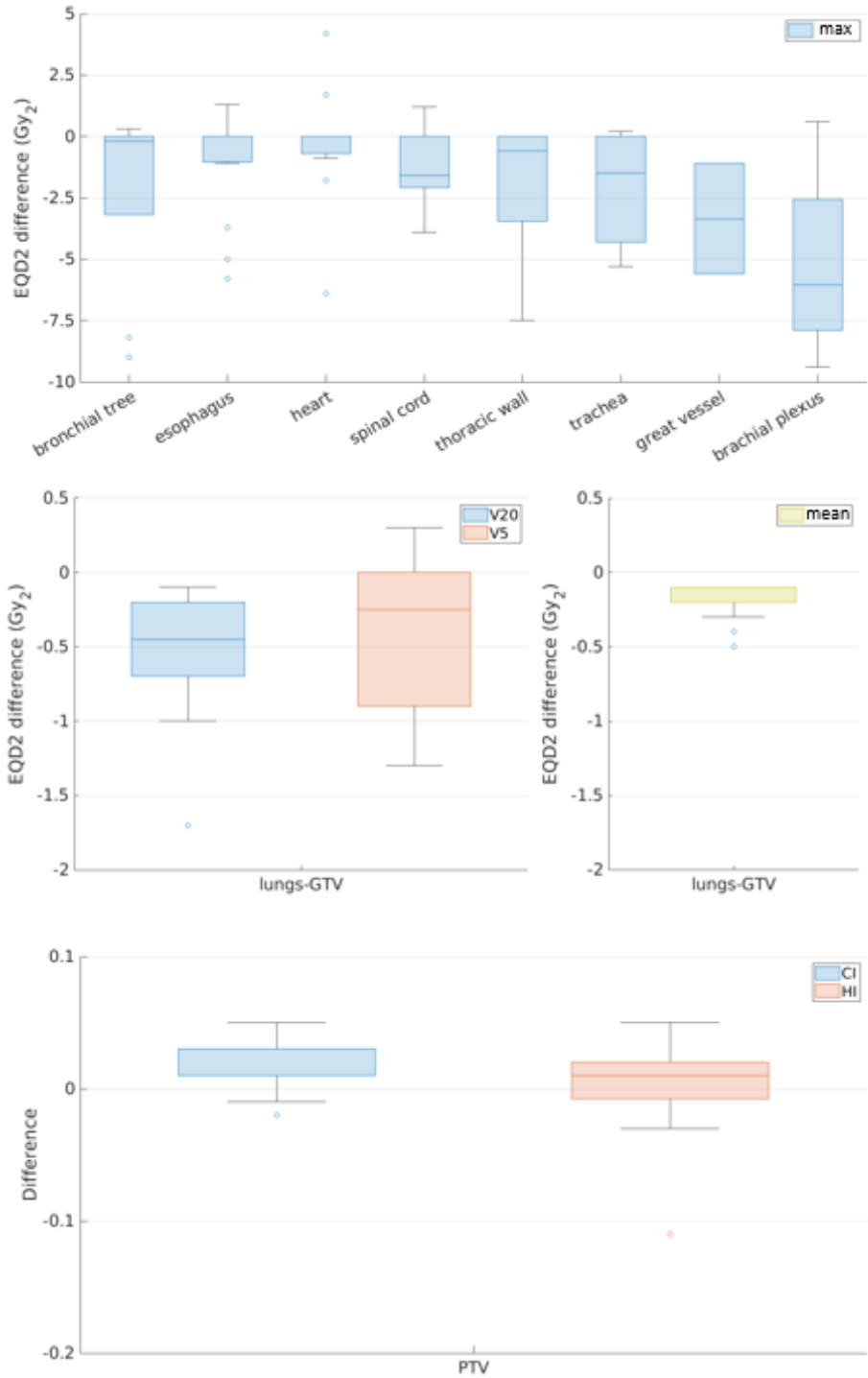

Figure 3: Comparison of the most relevant plan parameters between the coplanar and non-coplanar reirradiation plans for all considered patients (boxplots report median values along with 25% and 75% percentiles). The conformity index in the PTV is given by $CI = \frac{(V_{PTV} \cap V_{d_{pres}})^2}{V_{PTV} V_{d_{pres}}}$ (where $V_{PTV}$ is the PTV volume and $V_{d_{pres}}$ is the total volume receiving the prescribed cumulative dose $d_{pres}$), while the homogeneity index is expressed as $HI = \frac{d_2 - d_{98}}{d_{pres}}$ (where $d_2$ and $d_{98}$ are the doses received by 2% and 98% of the PTV volume in the reirradiation plan, respectively).

# Supplementary material A  Optimization parameters

In this section, the mathematical formulation of the planning objectives and priorities used in the treatment plan optimization problem are provided for each individual patient, and the dose calculation algorithm which has been used for computing the dose-influence matrices is discussed. The following notation is used:

- $PTV$ = set of voxels belonging to the PTV
- $L$ = set of voxels belonging to the healthy lungs (i.e. excluding the GTV voxels)
- $BT$ = set of voxels belonging to the bronchial tree
- $BT$ = set of voxels belonging to the bronchial tree
- $TW$ = set of voxels belonging to the thoracic wall
- $E$ = set of voxels belonging to the esophagus
- $H$ = set of voxels belonging to the heart
- $SC$ = set of voxels belonging to the spinal cord
- $T$ = set of voxels belonging to the trachea
- $GV$ = set of voxels belonging to the great vessels
- $BPR$ = set of voxels belonging to the right brachial plexus
- $BPL$ = set of voxels belonging to the left brachial plexus
- $TH$ = set of voxels belonging to the thyroid
- $LV$ = set of voxels belonging to the healthy liver (i.e. excluding the GTV voxels)
- $NT$ = set of voxels belonging to the normal tissue (i.e. the entire body except for the PTV)

## A.1 Patient 1

For patient 1, the objective function in Equation (3) reads as follows:

$$f(\mathbf{e}) = \frac{1}{|PTV|} \sum_{i \in PTV} [20(58.4 - e_i)_+^2 + 10(e_i - 65)_+^2] \qquad (S1.1)$$

$$+ \frac{1}{|L|} \sum_{i \in L} \left[50\tilde{e}_i + 2000 \frac{1}{1+e^{-(\tilde{e}_i-20)/0.5}} + 2000 \frac{1}{1+e^{-(\tilde{e}_i-5)/0.5}}\right] \qquad (S1.2)$$

$$+ \frac{1}{|BT|} \sum_{i \in BT} 50\tilde{e}_i \qquad (S1.3)$$

$$+ \frac{1}{|TW|} \sum_{i \in TW} [10\tilde{e}_i + 0.1(\tilde{e}_i - 74.3)_+^2] \qquad (S1.4)$$

$$+ \frac{1}{|E|} \sum_{i \in E} [10\tilde{e}_i + 0.1(\tilde{e}_i - 114.2)_+^2] \qquad (S1.5)$$

$$+ \frac{1}{|H|} \sum_{i \in H} [10\tilde{e}_i + 0.1(\tilde{e}_i - 32.8)_+^2] \qquad (S1.6)$$

$$+ \frac{1}{|SC|} \sum_{i \in SC} [10\tilde{e}_i + 0.1(\tilde{e}_i - 30.0)_+^2] \qquad (S1.7)$$

$$+ \frac{1}{|T|} \sum_{i \in T} [10\tilde{e}_i + 0.1(\tilde{e}_i - 105.4)_+^2] \qquad (S1.8)$$

$$+ \frac{1}{|NT|} \sum_{i \in NT} 250(e_i - e_i^{max})_+^2 \qquad (S1.9)$$

The planning objectives in Equation (S1.2) are used to control the mean EQD2 in the lungs, as well as the volume of healthy lungs exposed to EQD2 values larger than 20 $Gy_2$ and 5 $Gy_2$, respectively. Different than classical dose-volume objectives, which are evaluated using the Heaviside step function $\Theta(x)$, the objective functions in Equation (S1.2) are evaluated using a continuously differentiable logistic sigmoid function $\zeta_\epsilon(x) = \frac{1}{1+e^{-(\tilde{e}_i - \tilde{e}_{ref})/\epsilon}}$ (where $\epsilon$ is called smoothness parameter [1]). For $\epsilon \neq 0$, a smooth approximation of a dose-volume objective can be defined with a non-vanishing gradient around $e$ = 20 $Gy_2$ and $e$ = 5 $Gy_2$, respectively. In this work, smoothness parameter was set to $\epsilon = 0.5$ (note that $\zeta_\epsilon(x) \to \Theta(x)$ for $\epsilon \to 0$). The planning objective in Equation (S1.9), instead, corresponds to the normal tissue objective (NTO) implemented in the Eclipse Treatment Planning System (Varian, A Siemens Healtineers Company), where $e_i^{max}$ is a voxel-dependent value defined as:

$$e_i^{max} = \begin{cases} e_0 & \text{, if } x_i < x_0 \\ e_0 e^{-\kappa(x_i - x_0)} + e_\infty \left(1 - e^{-\kappa(x_i - x_0)}\right) & \text{, if } x_i \geq x_0 \end{cases}$$

Here, $x_i$ indicates the distance of a normal tissue voxel $i$ from the PTV edge and the NTO parameters were set to $e_0 = 60.0$ Gy$_2$, $e_\infty = 12.0$ Gy$_2$, $x_0 = 0.0$ cm and $\kappa = 1.0$ cm$^{-1}$. Note that the planning objectives in Equations (S1.1) and (S1.9) are evaluated only for the EQD2 of the reirradiation plan and not the cumulative EQD2.

## A.2 Patient 2

For patient 2, the objective function in Equation (3) reads as follows:

$$f(\mathbf{e}) = \frac{\omega_1}{|PTV|} \sum_{i \in PTV} [20(63.5 - e_i)_+^2 + 10(e_i - 70.7)_+^2] \qquad (S2.1)$$

$$+ \frac{1}{|L|} \sum_{i \in L} \left[ 50\tilde{e}_i + 2000 \frac{1}{1 + e^{-(\tilde{e}_i - 20)/0.5}} + 2000 \frac{1}{1 + e^{-(\tilde{e}_i - 5)/0.5}} \right] \qquad (S2.2)$$

$$+ \frac{1}{|E|} \sum_{i \in E} [20\tilde{e}_i + 0.01(\tilde{e}_i - 17.4)_+^2] \qquad (S2.3)$$

$$+ \frac{1}{|H|} \sum_{i \in H} [10\tilde{e}_i + 10(\tilde{e}_i - 5.3)_+^2] \qquad (S2.4)$$

$$+ \frac{1}{|SC|} \sum_{i \in SC} [10\tilde{e}_i + 70(\tilde{e}_i - 7.1)_+^2] \qquad (S2.5)$$

$$+ \frac{\omega_6}{|NT|} \sum_{i \in NT} 250(e_i - e_i^{max})_+^2 \qquad (S2.6)$$

where the NTO parameters were set to $e_0 = 64.0$ Gy$_2$, $e_\infty = 12.8$ Gy$_2$, $x_0 = 0.0$ cm and $\kappa = 1.0$ cm$^{-1}$. To achieve a similar target coverage and dose conformity in both the coplanar and non-coplanar reirradiation plan, different priorities have been set for the planning objectives in Equations (S2.1) and (S2.6): $\omega_1 = 2.5$ and $\omega_6 = 2$ were set or the coplanar reirradiation plan, and $\omega_1 = 1$ and $\omega_6 = 1$ were set or the non-coplanar reirradiation plan.

## A.3 Patient 3

For patient 3, the objective function in Equation (3) reads as follows:

$$f(e) = \frac{\omega_1}{|PTV|} \sum_{i \in PTV} [20(31.25 - e_i)_+^2 + 10(e_i - 54.7)_+^2] \quad (S3.1)$$

$$+ \frac{1}{|L|} \sum_{i \in L} \left[ 50\tilde{e}_i + 2000 \frac{1}{1 + e^{-(\tilde{e}_i - 20)/0.5}} + 2000 \frac{1}{1 + e^{-(\tilde{e}_i - 5)/0.5}} \right] \quad (S3.2)$$

$$+ \frac{1}{|E|} \sum_{i \in E} [\tilde{e}_i + (\tilde{e}_i - 88.9)_+^2] \quad (S3.3)$$

$$+ \frac{1}{|H|} \sum_{i \in H} \tilde{e}_i \quad (S3.4)$$

$$+ \frac{1}{|SC|} \sum_{i \in SC} [\tilde{e}_i + (\tilde{e}_i - 38.4)_+^2] \quad (S3.5)$$

$$+ \frac{\omega_6}{|NT|} \sum_{i \in NT} 250(e_i - e_i^{max})_+^2 \quad (S3.6)$$

where the NTO parameters were set to $e_0 = 36.0$ Gy$_2$, $e_\infty = 7.2$ Gy$_2$, $x_0 = 0.0$ cm and $\kappa = 1.0$ cm$^{-1}$. Different priorities have been set for the planning objectives in Equations (S3.1) and (S3.6): $\omega_1 = 2.5$ and $\omega_6 = 2$ were set or the coplanar reirradiation plan, and $\omega_1 = 1$ and $\omega_6 = 1$ were set or the non-coplanar reirradiation plan.

## A.4 Patient 4

For patient 4, the objective function in Equation (3) reads as follows:

$$f(\mathbf{e}) = \frac{\omega_1}{|PTV|} \sum_{i \in PTV} [20(54.4 - e_i)_+^2 + 10(e_i - 70.2)_+^2] \quad (S4.1)$$

$$+ \frac{1}{|L|} \sum_{i \in L} \left[50\tilde{e}_i + 2000\frac{1}{1+e^{-(\tilde{e}_i-20)/0.5}} + 2000\frac{1}{1+e^{-(\tilde{e}_i-5)/0.5}}\right] \quad (S4.2)$$

$$+ \frac{1}{|BT|} \sum_{i \in BT} [10\tilde{e}_i + 20(\tilde{e}_i - 108.0)_+^2] \quad (S4.3)$$

$$+ \frac{1}{|E|} \sum_{i \in E} [10\tilde{e}_i + 20(\tilde{e}_i - 77.7)_+^2] \quad (S4.4)$$

$$+ \frac{1}{|H|} \sum_{i \in H} [\tilde{e}_i + 0.1(\tilde{e}_i - 39.9)_+^2] \quad (S4.5)$$

$$+ \frac{1}{|SC|} \sum_{i \in SC} [\tilde{e}_i + 0.1(\tilde{e}_i - 31.6)_+^2] \quad (S4.6)$$

$$+ \frac{1}{|T|} \sum_{i \in T} [10\tilde{e}_i + 20(\tilde{e}_i - 105.7)_+^2] \quad (S4.7)$$

$$+ \frac{1}{|GV|} \sum_{i \in GV} [10\tilde{e}_i + 20(\tilde{e}_i - 103.4)_+^2] \quad (S4.8)$$

$$+ \frac{\omega_9}{|NT|} \sum_{i \in NT} 250(e_i - e_i^{max})_+^2 \quad (S4.9)$$

where the NTO parameters were set to $e_0 = 67.5$ Gy$_2$, $e_\infty = 13.5$ Gy$_2$, $x_0 = 0.0$ cm and $\kappa = 1.0$ cm$^{-1}$. Different priorities have been set for the planning objectives in Equations (S4.1) and (S4.9): $\omega_1 = 2.5$ and $\omega_9 = 2$ were set or the coplanar reirradiation plan, and $\omega_1 = 1$ and $\omega_9 = 1$ were set or the non-coplanar reirradiation plan.

## A.5 Patient 5

For patient 5, the objective function in Equation (3) reads as follows:

$$f(\boldsymbol{e}) = \frac{1}{|PTV|} \sum_{i \in PTV} [20(49.6 - e_i)_+^2 + 10(e_i - 83.3)_+^2] \tag{S5.1}$$

$$+ \quad \frac{1}{|L|} \sum_{i \in L} \left[ 50\tilde{e}_i + 2000 \frac{1}{1 + e^{-(\tilde{e}_i - 20)/0.5}} + 2000 \frac{1}{1 + e^{-(\tilde{e}_i - 5)/0.5}} \right] \tag{S5.2}$$

$$+ \quad \frac{1}{|BT|} \sum_{i \in BT} \tilde{e}_i \tag{S5.3}$$

$$+ \quad \frac{1}{|E|} \sum_{i \in E} [20\tilde{e}_i + (\tilde{e}_i - 62.0)_+^2] \tag{S5.4}$$

$$+ \quad \frac{1}{|H|} \sum_{i \in H} \tilde{e}_i \tag{S5.5}$$

$$+ \quad \frac{1}{|SC|} \sum_{i \in SC} [20\tilde{e}_i + (\tilde{e}_i - 48.8)_+^2] \tag{S5.6}$$

$$+ \quad \frac{1}{|T|} \sum_{i \in T} [20\tilde{e}_i + (\tilde{e}_i - 60.0)_+^2] \tag{S5.7}$$

$$+ \quad \frac{1}{|NT|} \sum_{i \in NT} 250(e_i - e_i^{max})_+^2 \tag{S5.8}$$

where the NTO parameters were set to $e_0 = 70.0$ Gy$_2$, $e_\infty = 14.0$ Gy$_2$, $x_0 = 0.0$ cm and $\kappa = 1.0$ cm$^{-1}$.

## A.6 Patient 6

For patient 6, the objective function in Equation (3) reads as follows:

$$f(\boldsymbol{e}) = \frac{\omega_1}{|PTV|} \sum_{i \in PTV} [20(21.85 - e_i)_+^2 + 10(e_i - 24.85)_+^2] \tag{S6.1}$$

$$+ \frac{1}{|L|} \sum_{i \in L} \left[ 50\tilde{e}_i + 2000 \frac{1}{1 + e^{-(\tilde{e}_i - 20)/0.5}} + 2000 \frac{1}{1 + e^{-(\tilde{e}_i - 5)/0.5}} \right] \tag{S6.2}$$

$$+ \frac{1}{|BT|} \sum_{i \in BT} 10\tilde{e}_i \tag{S6.3}$$

$$+ \frac{1}{|E|} \sum_{i \in E} 10\tilde{e}_i \tag{S6.4}$$

$$+ \frac{1}{|H|} \sum_{i \in H} 10\tilde{e}_i \tag{S6.5}$$

$$+ \frac{1}{|SC|} \sum_{i \in SC} [10\tilde{e}_i + 10(\tilde{e}_i - 23.9)_+^2] \tag{S6.6}$$

$$+ \frac{1}{|T|} \sum_{i \in T} 10\tilde{e}_i \tag{S6.7}$$

$$+ \frac{1}{|GV|} \sum_{i \in GV} 10\tilde{e}_i \tag{S6.8}$$

$$+ \frac{\omega_9}{|NT|} \sum_{i \in NT} 250(e_i - e_i^{max})_+^2 \tag{S6.9}$$

where the NTO parameters were set to $e_0 = 25.8$ Gy$_2$, $e_\infty = 5.7$ Gy$_2$, $x_0 = 0.0$ cm and $\kappa = 1.0$ cm$^{-1}$. Different priorities have been set for the planning objectives in Equations (S6.1) and (S6.9): $\omega_1 = 2.5$ and $\omega_9 = 2$ were set or the coplanar reirradiation plan, and $\omega_1 = 1$ and $\omega_9 = 1$ were set or the non-coplanar reirradiation plan.

## A.7 Patient 7

For patient 7, the objective function in Equation (3) reads as follows:

$$f(\boldsymbol{e}) = \frac{\omega_1}{|PTV|} \sum_{i \in PTV} [20(83.3 - e_i)_+^2 + 10(e_i - 163.0)_+^2] \tag{S7.1}$$

$$+ \frac{1}{|L|} \sum_{i \in L} \left[ 50\tilde{e}_i + 2000 \frac{1}{1 + e^{-(\tilde{e}_i - 20)/0.5}} + 2000 \frac{1}{1 + e^{-(\tilde{e}_i - 5)/0.5}} \right] \tag{S7.2}$$

$$+ \frac{1}{|TW|} \sum_{i \in TW} [10\tilde{e}_i + 10(\tilde{e}_i - 24.4)_+^2] \tag{S7.3}$$

$$+ \frac{1}{|E|} \sum_{i \in E} [10\tilde{e}_i + 10(\tilde{e}_i - 25.4)_+^2] \tag{S7.4}$$

$$+ \frac{1}{|H|} \sum_{i \in H} [10\tilde{e}_i + 10(\tilde{e}_i - 64.8)_+^2] \tag{S7.5}$$

$$+ \frac{1}{|SC|} \sum_{i \in SC} [10\tilde{e}_i + 10(\tilde{e}_i - 8.5)_+^2] \tag{S7.6}$$

$$+ \frac{1}{|T|} \sum_{i \in T} [10\tilde{e}_i + 10(\tilde{e}_i - 13.2)_+^2] \tag{S7.7}$$

$$+ \frac{\omega_8}{|NT|} \sum_{i \in NT} 250(e_i - e_i^{max})_+^2 \tag{S7.8}$$

where the NTO parameters were set to $e_0 = 130.0 \text{ Gy}_2$, $e_\infty = 26.0 \text{ Gy}_2$, $x_0 = 0.0$ cm and $\kappa = 1.0$ cm$^{-1}$. Different priorities have been set for the planning objectives in Equations (S7.1) and (S7.8): $\omega_1 = 2.5$ and $\omega_8 = 2$ were set or the coplanar reirradiation plan, and $\omega_1 = 1$ and $\omega_8 = 1$ were set or the non-coplanar reirradiation plan.

## A.8 Patient 8

For patient 8, the objective function in Equation (3) reads as follows:

$$f(\boldsymbol{e}) = \frac{1}{|PTV|} \sum_{i \in PTV} [20(50.0 - e_i)_+^2 + 10(e_i - 83.3)_+^2] \quad (S8.1)$$

$$+ \frac{1}{|L|} \sum_{i \in L} \left[ 50\tilde{e}_i + 2000 \frac{1}{1 + e^{-(\tilde{e}_i - 20)/0.5}} + 2000 \frac{1}{1 + e^{-(\tilde{e}_i - 5)/0.5}} \right] \quad (S8.2)$$

$$+ \frac{1}{|BT|} \sum_{i \in BT} \tilde{e}_i \quad (S8.3)$$

$$+ \frac{1}{|TW|} \sum_{i \in TW} [10\tilde{e}_i + 10(\tilde{e}_i - 214.0)_+^2] \quad (S8.4)$$

$$+ \frac{1}{|E|} \sum_{i \in E} \tilde{e}_i \quad (S8.5)$$

$$+ \frac{1}{|H|} \sum_{i \in H} \tilde{e}_i \quad (S8.6)$$

$$+ \frac{1}{|SC|} \sum_{i \in SC} [10\tilde{e}_i + 10(\tilde{e}_i - 11.5)_+^2] \quad (S8.7)$$

$$+ \frac{1}{|T|} \sum_{i \in T} [\tilde{e}_i + 0.1(\tilde{e}_i - 39.9)_+^2] \quad (S8.8)$$

$$+ \frac{1}{|NT|} \sum_{i \in NT} 250(e_i - e_i^{max})_+^2 \quad (S8.9)$$

where the NTO parameters were set to $e_0 = 64.0$ Gy$_2$, $e_\infty = 12.8$ Gy$_2$, $x_0 = 0.0$ cm and $\kappa = 1.0$ cm$^{-1}$.

## A.9 Patient 9

For patient 9, the objective function in Equation (3) reads as follows:

$$f(\mathbf{e}) \;=\; \frac{\omega_1}{|PTV|} \sum_{i \in PTV} [20(58.8 - e_i)_+^2 + 10(e_i - 61.2)_+^2] \tag{S9.1}$$

$$+ \;\; \frac{1}{|L|} \sum_{i \in L} \left[ 50\tilde{e}_i + 2000 \frac{1}{1 + e^{-(\tilde{e}_i - 20)/0.5}} + 2000 \frac{1}{1 + e^{-(\tilde{e}_i - 5)/0.5}} \right] \tag{S9.2}$$

$$+ \;\; \frac{1}{|E|} \sum_{i \in E} [10\tilde{e}_i + 10(\tilde{e}_i - 75.2)_+^2] \tag{S9.3}$$

$$+ \;\; \frac{1}{|H|} \sum_{i \in H} 0.1\tilde{e}_i \tag{S9.4}$$

$$+ \;\; \frac{1}{|SC|} \sum_{i \in SC} [10\tilde{e}_i + 10(\tilde{e}_i - 29.9)_+^2] \tag{S9.5}$$

$$+ \;\; \frac{1}{|T|} \sum_{i \in T} [20\tilde{e}_i + 10(\tilde{e}_i - 82.0)_+^2] \tag{S9.6}$$

$$+ \;\; \frac{1}{|BPR|} \sum_{i \in BPR} 10\tilde{e}_i \tag{S9.7}$$

$$+ \;\; \frac{\omega_8}{|NT|} \sum_{i \in NT} 250(e_i - e_i^{max})_+^2 \tag{S9.8}$$

where the NTO parameters were set to $e_0 = 64.0$ Gy$_2$, $e_\infty = 12.8$ Gy$_2$, $x_0 = 0.0$ cm and $\kappa = 1.0$ cm$^{-1}$. Different priorities have been set for the planning objectives in Equations (S9.1) and (S9.8): $\omega_1 = 2.5$ and $\omega_8 = 2$ were set or the coplanar reirradiation plan, and $\omega_1 = 1$ and $\omega_8 = 1$ were set or the non-coplanar reirradiation plan.

## A.10 Patient 10

For patient 10, the objective function in Equation (3) reads as follows:

$$f(e) = \frac{\omega_1}{|PTV|} \sum_{i \in PTV} [20(58.8 - e_i)_+^2 + 10(e_i - 61.2)_+^2] \quad (S10.1)$$

$$+ \frac{1}{|L|} \sum_{i \in L} \left[50\tilde{e}_i + 2000 \frac{1}{1 + e^{-(\tilde{e}_i - 20)/0.5}} + 2000 \frac{1}{1 + e^{-(\tilde{e}_i - 5)/0.5}}\right] \quad (S10.2)$$

$$+ \frac{1}{|BT|} \sum_{i \in BT} 20\tilde{e}_i \quad (S10.3)$$

$$+ \frac{1}{|TW|} \sum_{i \in TW} 0.1\tilde{e}_i \quad (S10.4)$$

$$+ \frac{1}{|E|} \sum_{i \in E} [10\tilde{e}_i + 10(\tilde{e}_i - 58.6)_+^2] \quad (S10.5)$$

$$+ \frac{1}{|H|} \sum_{i \in H} 0.1\tilde{e}_i \quad (S10.6)$$

$$+ \frac{1}{|SC|} \sum_{i \in SC} \tilde{e}_i \quad (S10.7)$$

$$+ \frac{1}{|T|} \sum_{i \in T} 20\tilde{e}_i \quad (S10.8)$$

$$+ \frac{\omega_8}{|NT|} \sum_{i \in NT} 250(e_i - e_i^{max})_+^2 \quad (S10.9)$$

where the NTO parameters were set to $e_0 = 58.8$ Gy$_2$, $e_\infty = 11.8$ Gy$_2$, $x_0 = 0.0$ cm and $\kappa = 1.0$ cm$^{-1}$. Different priorities have been set for the planning objectives in Equations (S10.1) and (S10.9): $\omega_1 = 2.5$ and $\omega_9 = 2$ were set or the coplanar reirradiation plan, and $\omega_1 = 1$ and $\omega_9 = 1$ were set or the non-coplanar reirradiation plan.

## A.11 Patient 11

For patient 11, the objective function in Equation (3) reads as follows:

$$f(\mathbf{e}) = \frac{\omega_1}{|PTV|} \sum_{i \in PTV} [20(52.1 - e_i)_+^2 + 10(e_i - 59.7)_+^2] \quad (S11.1)$$

$$+ \frac{1}{|E|} \sum_{i \in E} [20\tilde{e}_i + 0.1(\tilde{e}_i - 62.5)_+^2] \quad (S11.2)$$

$$+ \frac{1}{|SC|} \sum_{i \in SC} [10\tilde{e}_i + 10(\tilde{e}_i - 17.1)_+^2] \quad (S11.3)$$

$$+ \frac{1}{|T|} \sum_{i \in T} [10\tilde{e}_i + (\tilde{e}_i - 62.9)_+^2] \quad (S11.4)$$

$$+ \frac{1}{|BPR|} \sum_{i \in BPR} 10\tilde{e}_i \quad (S11.5)$$

$$+ \frac{1}{|BPL|} \sum_{i \in BPL} 10\tilde{e}_i \quad (S11.6)$$

$$+ \frac{1}{|TH|} \sum_{i \in TH} [10\tilde{e}_i + 0.1(\tilde{e}_i - 0.4)_+^2] \quad (S11.7)$$

$$+ \frac{\omega_8}{|NT|} \sum_{i \in NT} 250(e_i - e_i^{max})_+^2 \quad (S11.8)$$

where the NTO parameters were set to $e_0 = 55.0$ Gy$_2$, $e_\infty = 11.0$ Gy$_2$, $x_0 = 0.0$ cm and $\kappa = 1.0$ cm$^{-1}$. Different priorities have been set for the planning objectives in Equations (S11.1) and (S11.8): $\omega_1 = 2.5$ and $\omega_8 = 2$ were set or the coplanar reirradiation plan, and $\omega_1 = 1$ and $\omega_8 = 1$ were set or the non-coplanar reirradiation plan.

## A.12 Patient 12

For patient 12, the objective function in Equation (3) reads as follows:

$$f(\mathbf{e}) = \frac{\omega_1}{|PTV|} \sum_{i \in PTV} [20(59.2 - e_i)_+^2 + 10(e_i - 60.8)_+^2] \tag{S12.1}$$

$$+ \frac{1}{|L|} \sum_{i \in L} \left[ 50\widetilde{e}_i + 2000 \frac{1}{1 + e^{-(\widetilde{e}_i - 20)/0.5}} + 2000 \frac{1}{1 + e^{-(\widetilde{e}_i - 5)/0.5}} \right] \tag{S12.2}$$

$$+ \frac{1}{|BT|} \sum_{i \in BT} [20\widetilde{e}_i + (\widetilde{e}_i - 52.3)_+^2] \tag{S12.3}$$

$$+ \frac{1}{|E|} \sum_{i \in E} [10\widetilde{e}_i + 10(\widetilde{e}_i - 52.3)_+^2] \tag{S12.4}$$

$$+ \frac{1}{|H|} \sum_{i \in H} [10\widetilde{e}_i + 10(\widetilde{e}_i - 51.6)_+^2] \tag{S12.5}$$

$$+ \frac{1}{|SC|} \sum_{i \in SC} [10\widetilde{e}_i + 10(\widetilde{e}_i - 23.8)_+^2] \tag{S12.6}$$

$$+ \frac{1}{|T|} \sum_{i \in T} [\widetilde{e}_i + 0.1(\widetilde{e}_i - 52.4)_+^2] \tag{S12.7}$$

$$+ \frac{1}{|LV|} \sum_{i \in LV} [20\widetilde{e}_i + 10(\widetilde{e}_i - 52.1)_+^2] \tag{S12.8}$$

$$+ \frac{\omega_9}{|NT|} \sum_{i \in NT} 250(e_i - e_i^{max})_+^2 \tag{S12.9}$$

where the NTO parameters were set to $e_0 = 36.0$ Gy$_2$, $e_\infty = 7.2$ Gy$_2$, $x_0 = 0.0$ cm and $\kappa = 1.0$ cm$^{-1}$. Different priorities have been set for the planning objectives in Equations (S12.1) and (S12.9): $\omega_1 = 2.5$ and $\omega_9 = 2$ were set or the coplanar reirradiation plan, and $\omega_1 = 1$ and $\omega_9 = 1$ were set or the non-coplanar reirradiation plan.

## A.13 Patient 13

For patient 13, the objective function in Equation (3) reads as follows:

$$f(\mathbf{e}) = \frac{1}{|PTV|} \sum_{i \in PTV} [20(52.8 - e_i)_+^2 + 10(e_i - 56.0)_+^2] \tag{S13.1}$$

$$+ \frac{1}{|L|} \sum_{i \in L} \left[50\tilde{e}_i + 2000 \frac{1}{1 + e^{-(\tilde{e}_i - 20)/0.5}} + 2000 \frac{1}{1 + e^{-(\tilde{e}_i - 5)/0.5}}\right] \tag{S13.2}$$

$$+ \frac{1}{|BT|} \sum_{i \in BT} [10\tilde{e}_i + 10(\tilde{e}_i - 115.3)_+^2] \tag{S13.3}$$

$$+ \frac{1}{|TW|} \sum_{i \in TW} [10\tilde{e}_i + 10(\tilde{e}_i - 60.8)_+^2] \tag{S13.4}$$

$$+ \frac{1}{|E|} \sum_{i \in E} [10\tilde{e}_i + 10(\tilde{e}_i - 71.3)_+^2] \tag{S13.5}$$

$$+ \frac{1}{|H|} \sum_{i \in H} [\tilde{e}_i + (\tilde{e}_i - 60.0)_+^2] \tag{S13.6}$$

$$+ \frac{1}{|SC|} \sum_{i \in SC} [10\tilde{e}_i + 10(\tilde{e}_i - 22.4)_+^2] \tag{S13.7}$$

$$+ \frac{1}{|T|} \sum_{i \in T} [10\tilde{e}_i + 10(\tilde{e}_i - 80.4)_+^2] \tag{S13.8}$$

$$+ \frac{1}{|BPR|} \sum_{i \in BPR} [\tilde{e}_i + (\tilde{e}_i - 6.2)_+^2] \tag{S13.9}$$

$$+ \frac{1}{|NT|} \sum_{i \in NT} 250(e_i - e_i^{max})_+^2 \tag{S13.10}$$

where the NTO parameters were set to $e_0 = 74.9 \text{ Gy}_2$, $e_\infty = 15.0 \text{ Gy}_2$, $x_0 = 0.0 \text{ cm}$ and $\kappa = 1.0 \text{ cm}^{-1}$.

## A.14 Patient 14

For patient 14, the objective function in Equation (3) reads as follows:

$$f(\mathbf{e}) = \frac{\omega_1}{|PTV|} \sum_{i \in PTV} [20(63.6 - e_i)_+^2 + 10(e_i - 64.9)_+^2] \tag{S14.1}$$

$$+ \frac{1}{|L|} \sum_{i \in L} \left[ 50\tilde{e}_i + 2000 \frac{1}{1 + e^{-(\tilde{e}_i - 20)/0.5}} + 2000 \frac{1}{1 + e^{-(\tilde{e}_i - 5)/0.5}} \right] \tag{S14.2}$$

$$+ \frac{1}{|E|} \sum_{i \in E} [10\tilde{e}_i + 10(\tilde{e}_i - 56.7)_+^2] \tag{S14.3}$$

$$+ \frac{1}{|H|} \sum_{i \in H} 20\tilde{e}_i \tag{S14.4}$$

$$+ \frac{1}{|SC|} \sum_{i \in SC} [10\tilde{e}_i + 10(\tilde{e}_i - 18.9)_+^2] \tag{S14.5}$$

$$+ \frac{\omega_6}{|NT|} \sum_{i \in NT} 250(e_i - e_i^{max})_+^2 \tag{S14.6}$$

where the NTO parameters were set to $e_0 = 68.7$ Gy$_2$, $e_\infty = 13.7$ Gy$_2$, $x_0 = 0.0$ cm and $\kappa = 1.0$ cm$^{-1}$. Different priorities have been set for the planning objectives in Equations (S14.1) and (S14.6): $\omega_1 = 2.5$ and $\omega_6 = 2$ were set or the coplanar reirradiation plan, and $\omega_1 = 1$ and $\omega_6 = 1$ were set or the non-coplanar reirradiation plan.

## A.15 Patient 15

For patient 15, the objective function in Equation (3) reads as follows:

$$f(e) = \frac{1}{|PTV|} \sum_{i \in PTV} [20(51.9 - e_i)_+^2 + 10(e_i - 53.2)_+^2] \qquad (S15.1)$$

$$+ \quad \frac{1}{|L|} \sum_{i \in L} \left[ 50\tilde{e}_i + 2000 \frac{1}{1 + e^{-(\tilde{e}_i - 20)/0.5}} + 2000 \frac{1}{1 + e^{-(\tilde{e}_i - 5)/0.5}} \right] \qquad (S15.2)$$

$$+ \quad \frac{1}{|BT|} \sum_{i \in BT} 10\tilde{e}_i \qquad (S15.3)$$

$$+ \quad \frac{1}{|E|} \sum_{i \in E} [10\tilde{e}_i + (\tilde{e}_i - 18.7)_+^2] \qquad (S15.4)$$

$$+ \quad \frac{1}{|H|} \sum_{i \in H} [20\tilde{e}_i + 20(\tilde{e}_i - 59.4)_+^2] \qquad (S15.5)$$

$$+ \quad \frac{1}{|SC|} \sum_{i \in SC} [10\tilde{e}_i + 10(\tilde{e}_i - 38.2)_+^2] \qquad (S15.6)$$

$$+ \quad \frac{1}{|T|} \sum_{i \in T} [\tilde{e}_i + 0.1(\tilde{e}_i - 58.8)_+^2] \qquad (S15.7)$$

$$+ \quad \frac{1}{|NT|} \sum_{i \in NT} 250(e_i - e_i^{max})_+^2 \qquad (S15.8)$$

where the NTO parameters were set to $e_0 = 56.1$ Gy$_2$, $e_\infty = 11.2$ Gy$_2$, $x_0 = 0.0$ cm and $\kappa = 1.0$ cm$^{-1}$.

## A.16 Dose calculation algorithm

Calculation of the dose-influence matrix is performed with the open-source radiotherapy planning research platform CERR [2], using a quadrant infinite beam (QIB) algorithm [3]. The bixel size is set to 5 x 5 mm$^2$ and the photon energy is 6 MV for all candidate beam orientations. For each patient, a non-uniform dose grid size is used throughout the body: the original voxel resolution is used in the PTV and close to the PTV, where a larger dose gradient is expected. At a distance between 2 cm and 4 cm from the PTV edge, medium-size voxels are used with 8-fold volume, whereas at distances larger than 4 cm from the PTV edge large-size voxels are used with 64-fold volume. As previously shown by Mueller *et al* [4], the use of a non-uniform dose grid size allows to considerably enhance the computational efficiency with negligible trade-offs on the plan accuracy. All dosimetric results, however, are evaluated based on the finest (small) dose grid size.

## References Supplementary material A

# Supplementary material B  Results

In this section, the dosimetric results and the dose distributions obtained using both the coplanar and non-coplanar reirradiation plans are reported for each individual patient.

## B.1 Patient 1

The dose distributions along with the corresponding gantry-couch paths obtained for patient 1 are reported in Figure S1 for both the coplanar and non-coplanar reirradiation plans. The dosimetric results are instead detailed in Table S1.

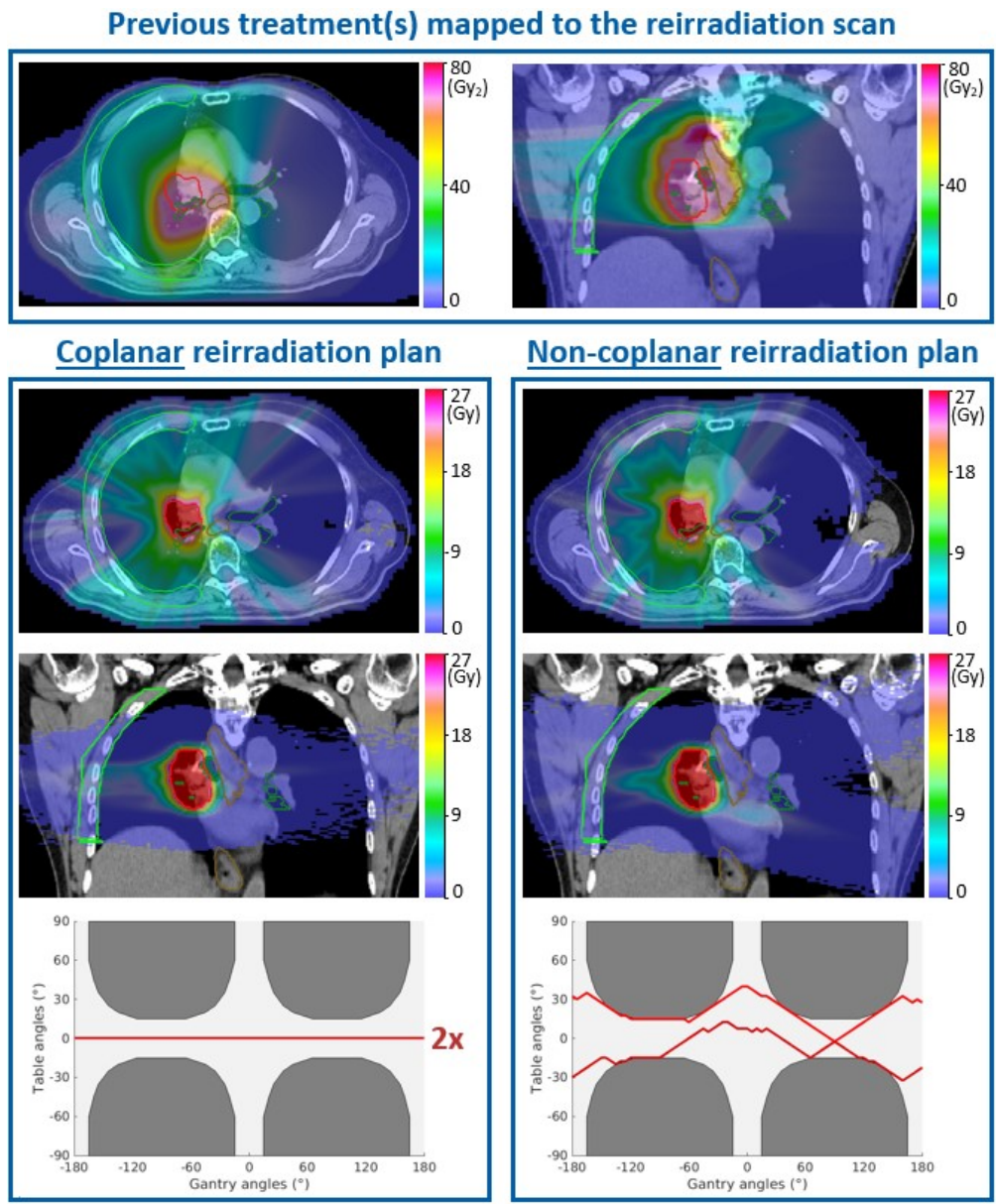

*Figure S1: Comparison of the coplanar and non-coplanar reirradiation plans generated for patient 1. Contours of PTV (red), bronchial tree (dark green), esophagus (brown) and thoracic wall (light green) are delineated in both the transversal and coronal planes of the reirradiation scan. The gantry-couch paths for both reirradiation plans are also shown, where dark grey regions indicate beam orientations leading to collision between gantry and couch (and are therefore excluded from the set of candidate beam orientations).*

Table S1: Cumulative EQD2 metrics for patient 1 achieved both prior to reirradiation and after reirradiation using the coplanar and non-coplanar plans. Percentage differences in dosimetric values between the coplanar and non-coplanar reirradiation plans are reported for each parameter. The conformity index in the PTV is given by $CI = \frac{(V_{PTV} \cap V_{d_{pres}})^2}{V_{PTV} V_{d_{pres}}}$ (where $V_{PTV}$ is the PTV volume and $V_{d_{pres}}$ is the total volume receiving the prescribed cumulative dose $d_{pres}$), while the homogeneity index is expressed as $HI = \frac{d_2 - d_{98}}{d_{pres}}$ (where $d_2$ and $d_{98}$ are the doses received by 2% and 98% of the PTV volume in the reirradiation plan, respectively).

| OAR | EQD2 parameter | Previous treatment(s) | Coplanar reirradiation plan | Non-coplanar reirradiation plan |
|---|---|---|---|---|
| Bronchial tree[#] | $D_{max}$ (Gy$_2$) | 71.4 | 124.0 | 122.5 (-1.2%) |
| | $D_{mean}$ (Gy$_2$) | 37.8 | 40.4 | 40.2 (-0.5%) |
| Esophagus | $D_{max}$ (Gy$_2$) | 126.9 | 126.9 | 126.9 (=) |
| | $D_{mean}$ (Gy$_2$) | 35.5 | 37.0 | 36.8 (-0.5%) |
| Heart | $D_{max}$ (Gy$_2$) | 36.5 | 45.0 | 49.2 (+9.3%) |
| | $D_{mean}$ (Gy$_2$) | 1.51 | 1.60 | 1.98 (+23.8%) |
| Spinal cord | $D_{max}$ (Gy$_2$) | 33.3 | 36.0 | 33.9 (-5.8%) |
| | $D_{mean}$ (Gy$_2$) | 9.99 | 11.0 | 10.9 (-0.9%) |
| Thoracic wall | $D_{max}$ (Gy$_2$) | 82.6 | 82.6 | 82.6 (=) |
| | $D_{mean}$ (Gy$_2$) | 12.6 | 15.1 | 14.9 (-1.3%) |
| Trachea | $D_{max}$ (Gy$_2$) | 117.2 | 117.2 | 117.2 (=) |
| | $D_{mean}$ (Gy$_2$) | 45.7 | 45.8 | 45.8 (=) |
| Lungs-GTV | $D_{mean}$ (Gy$_2$) | 11.7 | 14.5 | 14.4 (-0.7%) |
| | $V_{5Gy2}$ (%) | 47.9 | 49.2 | 48.9 (-0.6%) |
| | $V_{20Gy2}$ (%) | 19.5 | 23.1 | 22.1 (-4.3%) |
| PTV | HI | - | 0.18 | 0.17 (-5.6%) |
| | CI | - | 0.73 | 0.73 (=) |

[#] OARs which overlap with the PTV

B.2 Patient 2

The dose distributions along with the corresponding gantry-couch paths obtained for patient 2 are reported in Figure S2 for both the coplanar and non-coplanar reirradiation plans. The dosimetric results are instead detailed in Table S2.

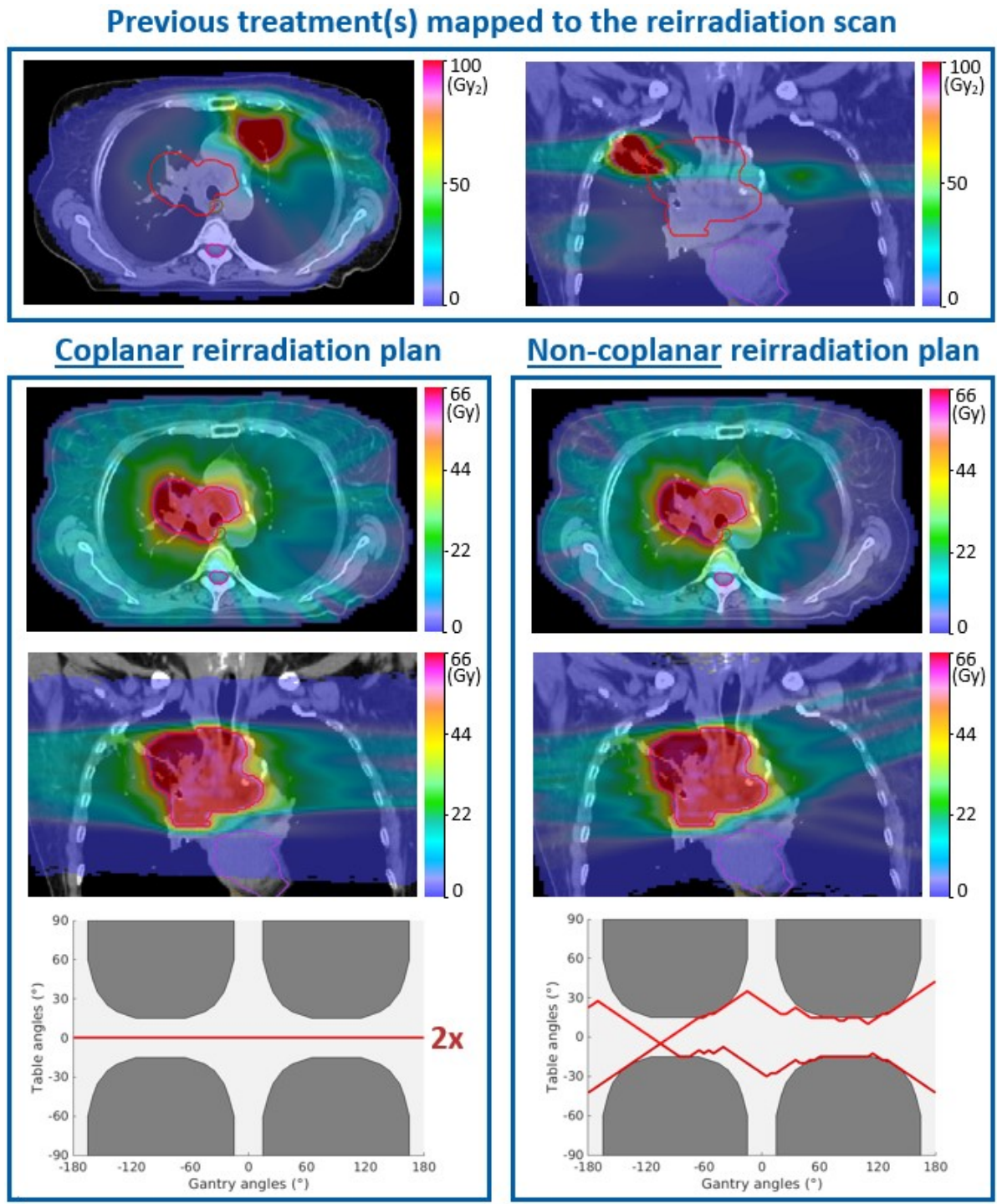

*Figure S2: Comparison of the coplanar and non-coplanar reirradiation plans generated for patient 2. Contours of PTV (red), spinal cord (pink), esophagus (brown) and heart (violet) are delineated in both the transversal and coronal planes of the reirradiation scan. The gantry-couch paths for both reirradiation plans are also shown, where dark grey regions indicate beam orientations leading to collision between gantry and couch (and are therefore excluded from the set of candidate beam orientations).*

*Table S2: Cumulative EQD2 metrics for patient 2 achieved both prior to reirradiation and after reirradiation using the coplanar and non-coplanar plans. Percentage differences in dosimetric values between the coplanar and non-coplanar reirradiation plans are reported for each parameter.*

| OAR | EQD2 parameter | Previous treatment(s) | Coplanar reirradiation plan | Non-coplanar reirradiation plan |
|---|---|---|---|---|
| Esophagus[#] | $D_{max}$ ($Gy_2$) | 19.3 | 63.8 | 58.0 (-9.1%) |
| | $D_{mean}$ ($Gy_2$) | 4.05 | 13.1 | 9.79 (-25.3%) |
| Heart | $D_{max}$ ($Gy_2$) | 5.89 | 18.3 | 11.9 (-35.0%) |
| | $D_{mean}$ ($Gy_2$) | 0.90 | 1.08 | 1.07 (-0.9%) |
| Spinal cord | $D_{max}$ ($Gy_2$) | 7.87 | 12.1 | 10.0 (-17.4%) |
| | $D_{mean}$ ($Gy_2$) | 3.12 | 5.41 | 4.85 (-10.4%) |
| Lungs-GTV | $D_{mean}$ ($Gy_2$) | 9.37 | 16.0 | 15.5 (-3.1%) |
| | $V_{5Gy2}$ (%) | 30.4 | 43.8 | 42.9 (-2.1%) |
| | $V_{20Gy2}$ (%) | 9.62 | 24.6 | 21.9 (-11.0%) |
| PTV | HI | - | 0.18 | 0.20 (+11.1%) |
| | CI | - | 0.31 | 0.32 (+3.2) |

[#] OARs which overlap with the PTV

B.3 Patient 3

The dose distributions along with the corresponding gantry-couch paths obtained for patient 3 are reported in Figure S3 for both the coplanar and non-coplanar reirradiation plans. The dosimetric results are instead detailed in Table S3.

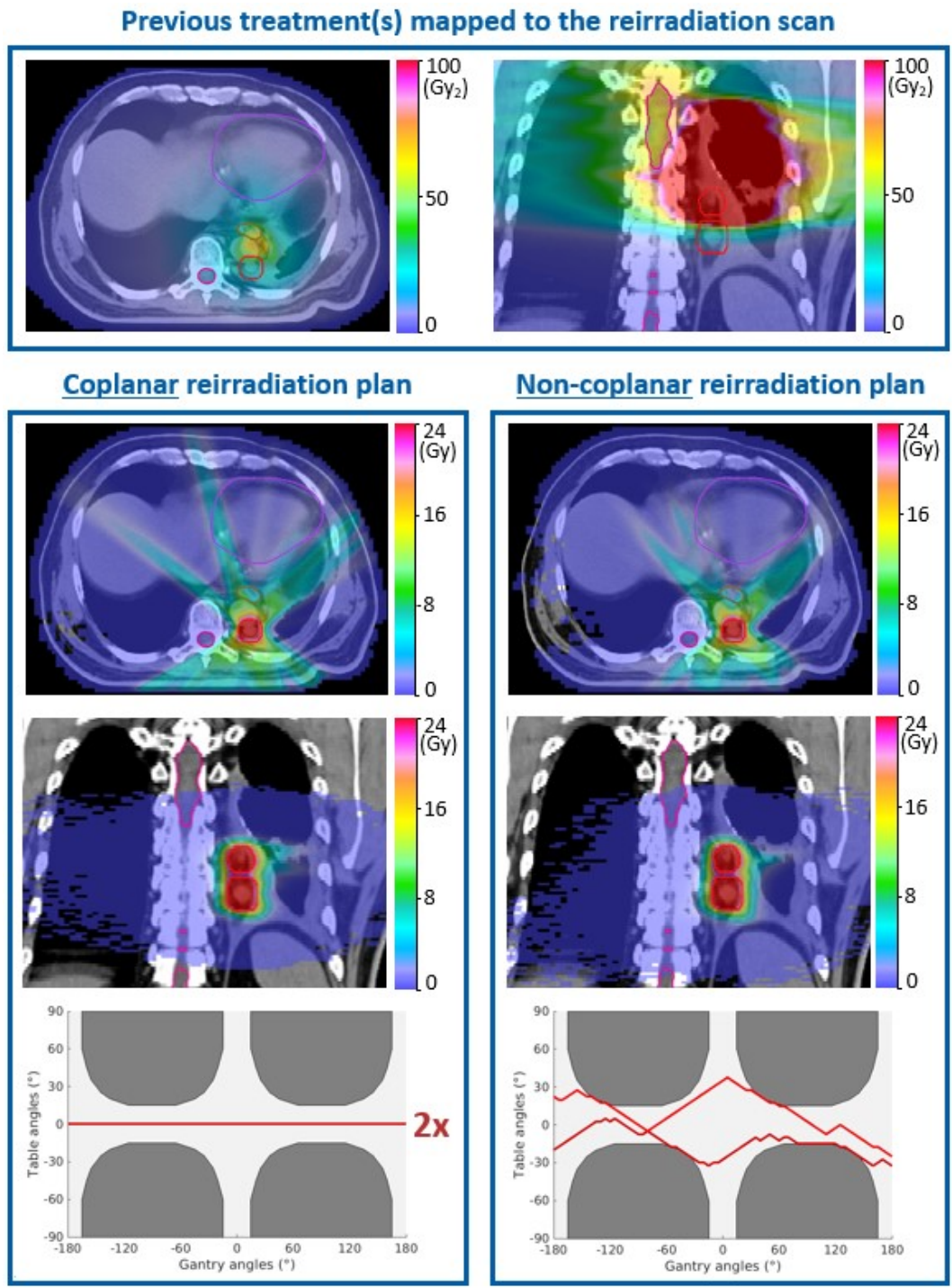

*Figure S3: Comparison of the coplanar and non-coplanar reirradiation plans generated for patient 3. Contours of PTV (red), spinal cord (pink), esophagus (brown) and heart (violet) are delineated in both the transversal and coronal planes of the reirradiation scan. The gantry-couch paths for both reirradiation plans are also shown, where dark grey regions indicate beam orientations leading to collision between gantry and couch (and are therefore excluded from the set of candidate beam orientations).*

*Table S3: Cumulative EQD2 metrics for patient 3 achieved both prior to reirradiation and after reirradiation using the coplanar and non-coplanar plans. Percentage differences in dosimetric values between the coplanar and non-coplanar reirradiation plans are reported for each parameter.*

| OAR | EQD2 parameter | Previous treatment(s) | Coplanar reirradiation plan | Non-coplanar reirradiation plan |
|---|---|---|---|---|
| Esophagus | $D_{max}$ ($Gy_2$) | 98.8 | 100.6 | 99.7 (-0.9%) |
| | $D_{mean}$ ($Gy_2$) | 36.7 | 38.1 | 37.5 (-1.6%) |
| Heart | $D_{max}$ ($Gy_2$) | 82.5 | 83.2 | 82.8 (-0.5%) |
| | $D_{mean}$ ($Gy_2$) | 22.6 | 23.4 | 23.0 (-1.7%) |
| Spinal cord | $D_{max}$ ($Gy_2$) | 42.7 | 42.8 | 42.8 (=) |
| | $D_{mean}$ ($Gy_2$) | 13.4 | 13.5 | 13.5 (=) |
| Lungs-GTV | $D_{mean}$ ($Gy_2$) | 18.1 | 18.2 | 18.1 (-0.5%) |
| | $V_{5Gy_2}$ (%) | 91.4 | 91.5 | 91.5 (=) |
| | $V_{20Gy_2}$ (%) | 29.3 | 29.4 | 29.3 (-0.3%) |
| PTV | HI | - | 0.42 | 0.42 (=) |
| | CI | - | 0.82 | 0.85 (+3.7) |

## B.4 Patient 4

The dose distributions along with the corresponding gantry-couch paths obtained for patient 4 are reported in Figure 3 of the main manuscript for both the coplanar and non-coplanar reirradiation plans. The dosimetric results are instead detailed in Table S4.

*Table S4: Cumulative EQD2 metrics for patient 4 achieved both prior to reirradiation and after reirradiation using the coplanar and non-coplanar plans. Percentage differences in dosimetric values between the coplanar and non-coplanar reirradiation plans are reported for each parameter.*

| OAR | EQD2 parameter | Previous treatment(s) | Coplanar reirradiation plan | Non-coplanar reirradiation plan |
| --- | --- | --- | --- | --- |
| Bronchial tree | $D_{max}$ (Gy$_2$) | 120.0 | 141.3 | 132.3 (-6.4%) |
|  | $D_{mean}$ (Gy$_2$) | 89.3 | 94.1 | 93.3 (-0.9%) |
| Esophagus | $D_{max}$ (Gy$_2$) | 86.3 | 95.5 | 90.9 (-5.2%) |
|  | $D_{mean}$ (Gy$_2$) | 41.5 | 42.1 | 41.9 (-0.5%) |
| Heart | $D_{max}$ (Gy$_2$) | 44.4 | 44.4 | 44.4 (=) |
|  | $D_{mean}$ (Gy$_2$) | 2.14 | 2.14 | 2.14 (=) |
| Spinal cord | $D_{max}$ (Gy$_2$) | 35.1 | 45.1 | 41.2 (-8.6%) |
|  | $D_{mean}$ (Gy$_2$) | 9.79 | 10.7 | 10.7 (=) |
| Great vessel | $D_{max}$ (Gy$_2$) | 114.9 | 120.6 | 115.0 (-4.6%) |
|  | $D_{mean}$ (Gy$_2$) | 42.8 | 43.7 | 43.5 (-0.5%) |
| Trachea | $D_{max}$ (Gy$_2$) | 117.5 | 127.6 | 122.7 (-3.8%) |
|  | $D_{mean}$ (Gy$_2$) | 51.7 | 52.7 | 52.5 (-0.4%) |
| Lungs-GTV | $D_{mean}$ (Gy$_2$) | 11.1 | 12.2 | 12.1 (-0.8%) |
|  | $V_{5Gy_2}$ (%) | 45.0 | 45.5 | 45.3 (-0.4%) |
|  | $V_{20Gy_2}$ (%) | 17.4 | 18.8 | 18.7 (-0.5%) |
| PTV | HI | - | 0.29 | 0.31 (+6.9%) |
|  | CI | - | 0.75 | 0.75 (=) |

B.5 Patient 5

The dose distributions along with the corresponding gantry-couch paths obtained for patient 5 are reported in Figure S4 for both the coplanar and non-coplanar reirradiation plans. The dosimetric results are instead detailed in Table S5.

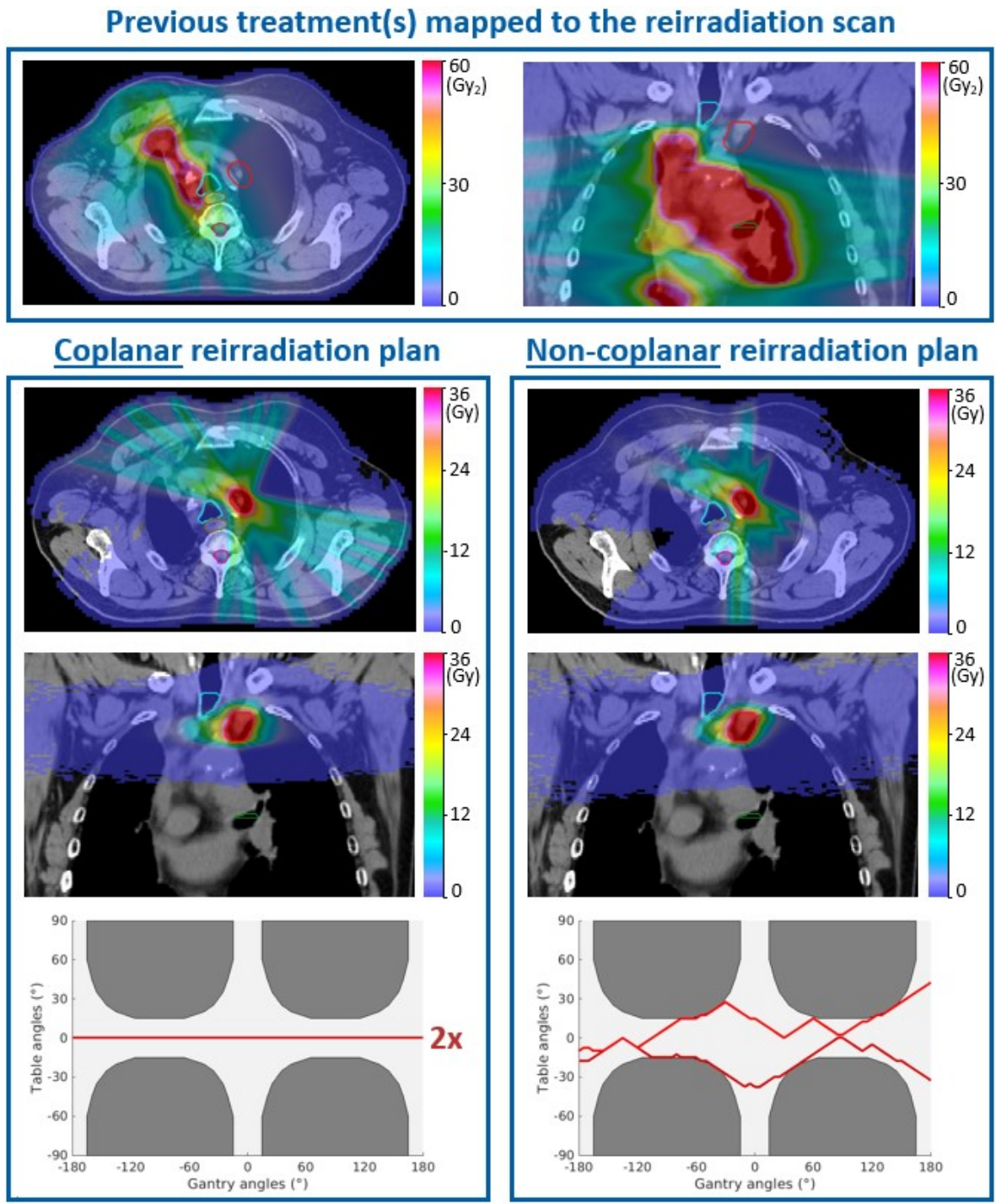

*Figure S4: Comparison of the coplanar and non-coplanar reirradiation plans generated for patient 5. Contours of PTV (red), spinal cord (pink), esophagus (brown) and trachea (light blue) are delineated in both the transversal and coronal planes of the reirradiation scan. The gantry-couch paths for both reirradiation plans are also shown, where dark grey regions indicate beam orientations leading to collision between gantry and couch (and are therefore excluded from the set of candidate beam orientations).*

*Table S5: Cumulative EQD2 metrics for patient 5 achieved both prior to reirradiation and after reirradiation using the coplanar and non-coplanar plans. Percentage differences in dosimetric values between the coplanar and non-coplanar reirradiation plans are reported for each parameter.*

| OAR | EQD2 parameter | Previous treatment(s) | Coplanar reirradiation plan | Non-coplanar reirradiation plan |
|---|---|---|---|---|
| Bronchial tree | $D_{max}$ (Gy$_2$) | 66.0 | 66.0 | 66.0 (=) |
|  | $D_{mean}$ (Gy$_2$) | 60.7 | 60.7 | 60.7 (=) |
| Esophagus | $D_{max}$ (Gy$_2$) | 68.8 | 68.8 | 68.8 (=) |
|  | $D_{mean}$ (Gy$_2$) | 32.2 | 32.3 | 32.4 (+0.3%) |
| Heart | $D_{max}$ (Gy$_2$) | 64.4 | 64.4 | 64.4 (=) |
|  | $D_{mean}$ (Gy$_2$) | 15.4 | 15.4 | 15.4 (=) |
| Spinal cord | $D_{max}$ (Gy$_2$) | 53.8 | 53.8 | 53.8 (=) |
|  | $D_{mean}$ (Gy$_2$) | 10.3 | 10.4 | 10.4 (=) |
| Trachea | $D_{max}$ (Gy$_2$) | 66.7 | 66.7 | 66.7 (=) |
|  | $D_{mean}$ (Gy$_2$) | 34.9 | 35.1 | 35.0 (-0.1%) |
| Lungs-GTV | $D_{mean}$ (Gy$_2$) | 17.5 | 17.9 | 17.8 (-0.6%) |
|  | $V_{5Gy2}$ (%) | 89.3 | 90.5 | 90.2 (-0.3%) |
|  | $V_{20Gy2}$ (%) | 27.9 | 28.7 | 28.5 (-0.7%) |
| PTV | HI | - | 0.44 | 0.44 (=) |
|  | CI | - | 0.88 | 0.89 (+1.1) |

## B.6 Patient 6

The dose distributions along with the corresponding gantry-couch paths obtained for patient 6 are reported in Figure S5 for both the coplanar and non-coplanar reirradiation plans. The dosimetric results are instead detailed in Table S6.

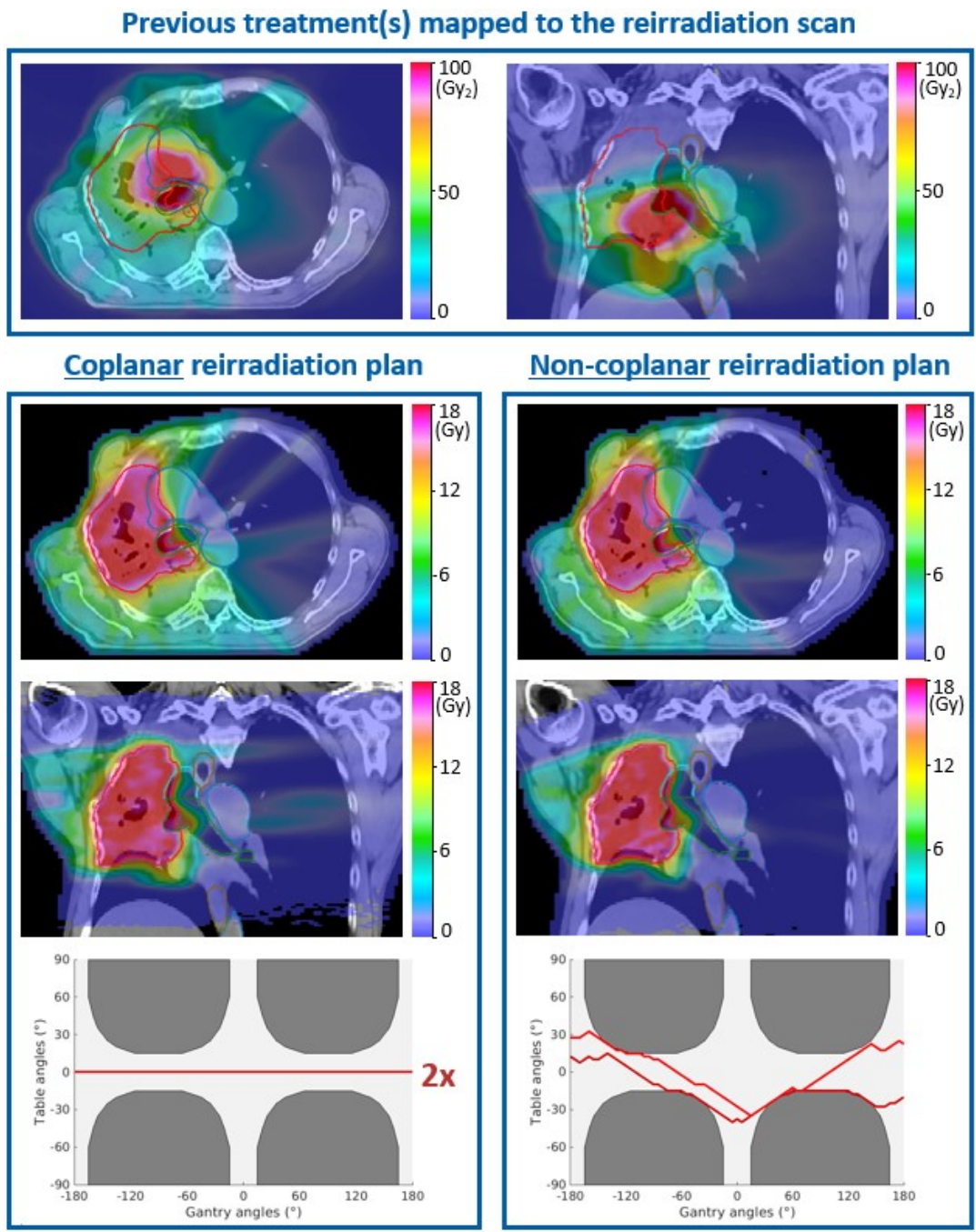

*Figure S5: Comparison of the coplanar and non-coplanar reirradiation plans generated for patient 6. Contours of PTV (red), bronchial tree (green), esophagus (brown) and great vessel (blue) are delineated in both the transversal and coronal planes of the reirradiation scan. The gantry-couch paths for both reirradiation plans are also shown, where dark grey regions indicate beam orientations leading to collision between gantry and couch (and are therefore excluded from the set of candidate beam orientations).*

*Table S6: Cumulative EQD2 metrics for patient 6 achieved both prior to reirradiation and after reirradiation using the coplanar and non-coplanar plans. Percentage differences in dosimetric values between the coplanar and non-coplanar reirradiation plans are reported for each parameter.*

| OAR | EQD2 parameter | Previous treatment(s) | Coplanar reirradiation plan | Non-coplanar reirradiation plan |
|---|---|---|---|---|
| Bronchial tree[#] | $D_{max}$ ($Gy_2$) | 131.3 | 142.4 | 142.7 (+0.2%) |
| | $D_{mean}$ ($Gy_2$) | 68.5 | 72.5 | 71.5 (-1.4%) |
| Esophagus[#] | $D_{max}$ ($Gy_2$) | 92.4 | 99.5 | 95.8 (-3.7%) |
| | $D_{mean}$ ($Gy_2$) | 19.6 | 20.6 | 20.0 (-2.9%) |
| Heart[#] | $D_{max}$ ($Gy_2$) | 144.7 | 159.2 | 157.4 (-1.1%) |
| | $D_{mean}$ ($Gy_2$) | 12.9 | 13.7 | 13.7 (=) |
| Spinal cord | $D_{max}$ ($Gy_2$) | 26.5 | 29.9 | 28.1 (-6.0%) |
| | $D_{mean}$ ($Gy_2$) | 2.69 | 3.65 | 3.54 (-3.0%) |
| Great vessel[#] | $D_{max}$ ($Gy_2$) | 143.5 | 159.6 | 158.5 (-0.7%) |
| | $D_{mean}$ ($Gy_2$) | 41.0 | 43.4 | 42.8 (-1.4%) |
| Trachea[#] | $D_{max}$ ($Gy_2$) | 49.6 | 52.7 | 50.0 (-5.1%) |
| | $D_{mean}$ ($Gy_2$) | 9.29 | 11.2 | 10.2 (-8.9%) |
| Lungs-GTV | $D_{mean}$ ($Gy_2$) | 9.22 | 10.1 | 9.88 (-2.2%) |
| | $V_{5Gy2}$ (%) | 53.5 | 54.3 | 53.9 (-0.7%) |
| | $V_{20Gy2}$ (%) | 14.4 | 14.9 | 14.4 (-0.5%) |
| PTV | HI | - | 0.22 | 0.27 (+18.5%) |
| | CI | - | 0.27 | 0.32 (+22.7%) |

[#] OARs which overlap with the PTV

B.7 Patient 7

The dose distributions along with the corresponding gantry-couch paths obtained for patient 7 are reported in Figure S6 for both the coplanar and non-coplanar reirradiation plans. The dosimetric results are instead detailed in Table S7.

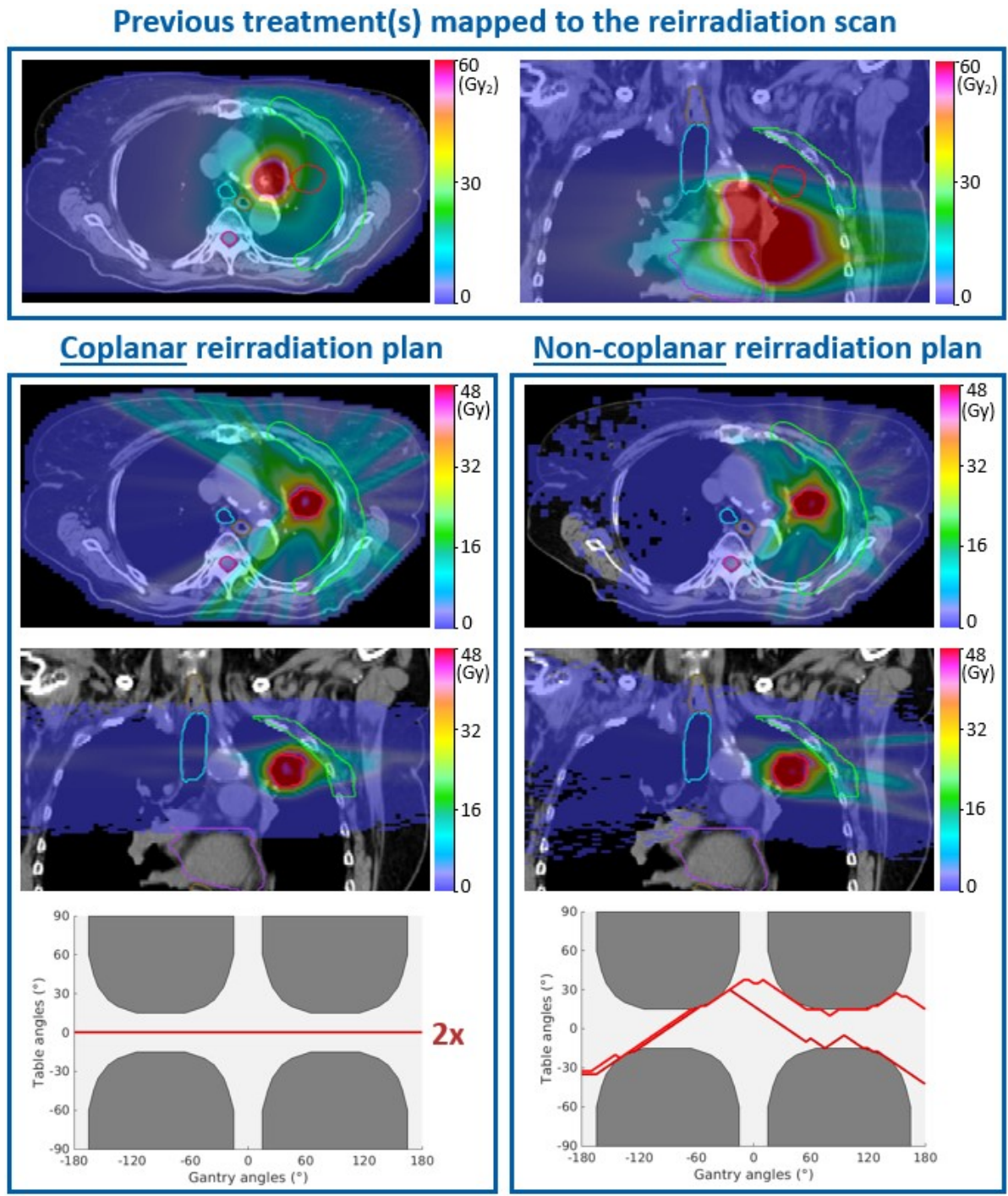

*Figure S6: Comparison of the coplanar and non-coplanar reirradiation plans generated for patient 7. Contours of PTV (red), thoracic wall (light green), trachea (light blue), heart (violet) and spinal cord (pink) are delineated in both the transversal and coronal planes of the reirradiation scan. The gantry-couch paths for both reirradiation plans are also shown, where dark grey regions indicate beam orientations leading to collision between gantry and couch (and are therefore excluded from the set of candidate beam orientations).*

*Table S7: Cumulative EQD2 metrics for patient 7 achieved both prior to reirradiation and after reirradiation using the coplanar and non-coplanar plans. Percentage differences in dosimetric values between the coplanar and non-coplanar reirradiation plans are reported for each parameter.*

| OAR | EQD2 parameter | Previous treatment(s) | Coplanar reirradiation plan | Non-coplanar reirradiation plan |
|---|---|---|---|---|
| Esophagus | $D_{max}$ (Gy$_2$) | 28.2 | 28.2 | 28.2 (=) |
|  | $D_{mean}$ (Gy$_2$) | 7.45 | 7.64 | 7.62 (-0.3%) |
| Heart | $D_{max}$ (Gy$_2$) | 72.0 | 72.0 | 72.0 (=) |
|  | $D_{mean}$ (Gy$_2$) | 14.2 | 14.2 | 14.2 (=) |
| Spinal cord | $D_{max}$ (Gy$_2$) | 9.46 | 10.9 | 9.47 (-13.1%) |
|  | $D_{mean}$ (Gy$_2$) | 1.77 | 1.93 | 1.96 (+1.6%) |
| Thoracic wall# | $D_{max}$ (Gy$_2$) | 27.1 | 98.7 | 91.2 (-7.6%) |
|  | $D_{mean}$ (Gy$_2$) | 5.61 | 14.2 | 12.4 (-12.7%) |
| Trachea | $D_{max}$ (Gy$_2$) | 14.7 | 15.2 | 14.8 (-2.6%) |
|  | $D_{mean}$ (Gy$_2$) | 2.12 | 2.47 | 2.27 (-8.1%) |
| Lungs-GTV | $D_{mean}$ (Gy$_2$) | 8.57 | 11.2 | 11.0 (-1.8%) |
|  | $V_{5Gy2}$ (%) | 32.7 | 35.9 | 34.6 (-3.6%) |
|  | $V_{20Gy2}$ (%) | 12.5 | 16.3 | 15.9 (-2.5%) |
| PTV | HI | - | 0.31 | 0.31 (=) |
|  | CI | - | 0.42 | 0.43 (+2.4%) |

# OARs which overlap with the PTV

B.8 Patient 8

The dose distributions along with the corresponding gantry-couch paths obtained for patient 8 are reported in Figure S7 for both the coplanar and non-coplanar reirradiation plans. The dosimetric results are instead detailed in Table S8.

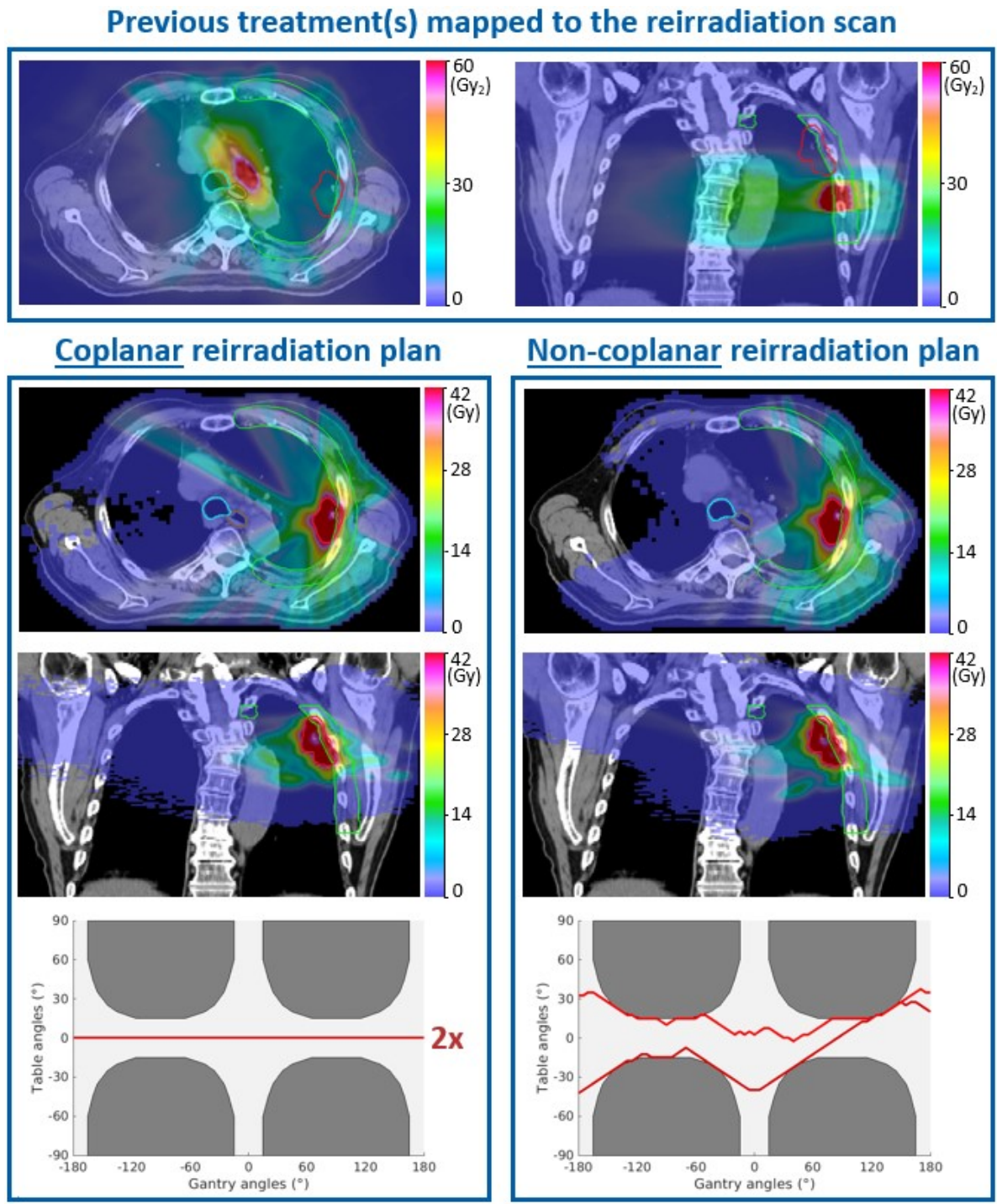

*Figure S7: Comparison of the coplanar and non-coplanar reirradiation plans generated for patient 8. Contours of PTV (red), thoracic wall (light green), trachea (light blue) and esophagus (brown) are delineated in both the transversal and coronal planes of the reirradiation scan. The gantry-couch paths for both reirradiation plans are also shown, where dark grey regions indicate beam orientations leading to collision between gantry and couch (and are therefore excluded from the set of candidate beam orientations).*

*Table S8: Cumulative EQD2 metrics for patient 8 achieved both prior to reirradiation and after reirradiation using the coplanar and non-coplanar plans. Percentage differences in dosimetric values between the coplanar and non-coplanar reirradiation plans are reported for each parameter.*

| OAR | EQD2 parameter | Previous treatment(s) | Coplanar reirradiation plan | Non-coplanar reirradiation plan |
|---|---|---|---|---|
| Bronchial tree | $D_{max}$ (Gy$_2$) | 56.9 | 56.9 | 56.9 (=) |
| | $D_{mean}$ (Gy$_2$) | 33.1 | 33.1 | 33.1 (=) |
| Esophagus | $D_{max}$ (Gy$_2$) | 45.8 | 45.8 | 45.8 (=) |
| | $D_{mean}$ (Gy$_2$) | 15.3 | 15.3 | 15.5 (+1.3%) |
| Heart | $D_{max}$ (Gy$_2$) | 64.1 | 64.1 | 64.1 (=) |
| | $D_{mean}$ (Gy$_2$) | 5.08 | 5.08 | 5.08 (=) |
| Spinal cord | $D_{max}$ (Gy$_2$) | 12.8 | 14.8 | 13.5 (-1.3%) |
| | $D_{mean}$ (Gy$_2$) | 1.95 | 2.21 | 2.12 (-0.1%) |
| Thoracic wall[#] | $D_{max}$ (Gy$_2$) | 237.8 | 270.9 | 270.3 (-0.2%) |
| | $D_{mean}$ (Gy$_2$) | 12.3 | 18.7 | 18.5 (-1.1%) |
| Trachea | $D_{max}$ (Gy$_2$) | 44.4 | 44.4 | 44.6 (+0.5%) |
| | $D_{mean}$ (Gy$_2$) | 12.5 | 12.6 | 12.8 (+1.6%) |
| Lungs-GTV | $D_{mean}$ (Gy$_2$) | 7.02 | 8.72 | 8.63 (-1.0%) |
| | $V_{5Gy2}$ (%) | 40.5 | 42.8 | 42.7 (-0.2%) |
| | $V_{20Gy2}$ (%) | 7.28 | 10.7 | 10.2 (-4.7%) |
| PTV | HI | - | 0.27 | 0.24 (-11.1%) |
| | CI | - | 0.53 | 0.53 (=) |

[#] OARs which overlap with the PTV

B.9 Patient 9

The dose distributions along with the corresponding gantry-couch paths obtained for patient 9 are reported in Figure S8 for both the coplanar and non-coplanar reirradiation plans. The dosimetric results are instead detailed in Table S9.

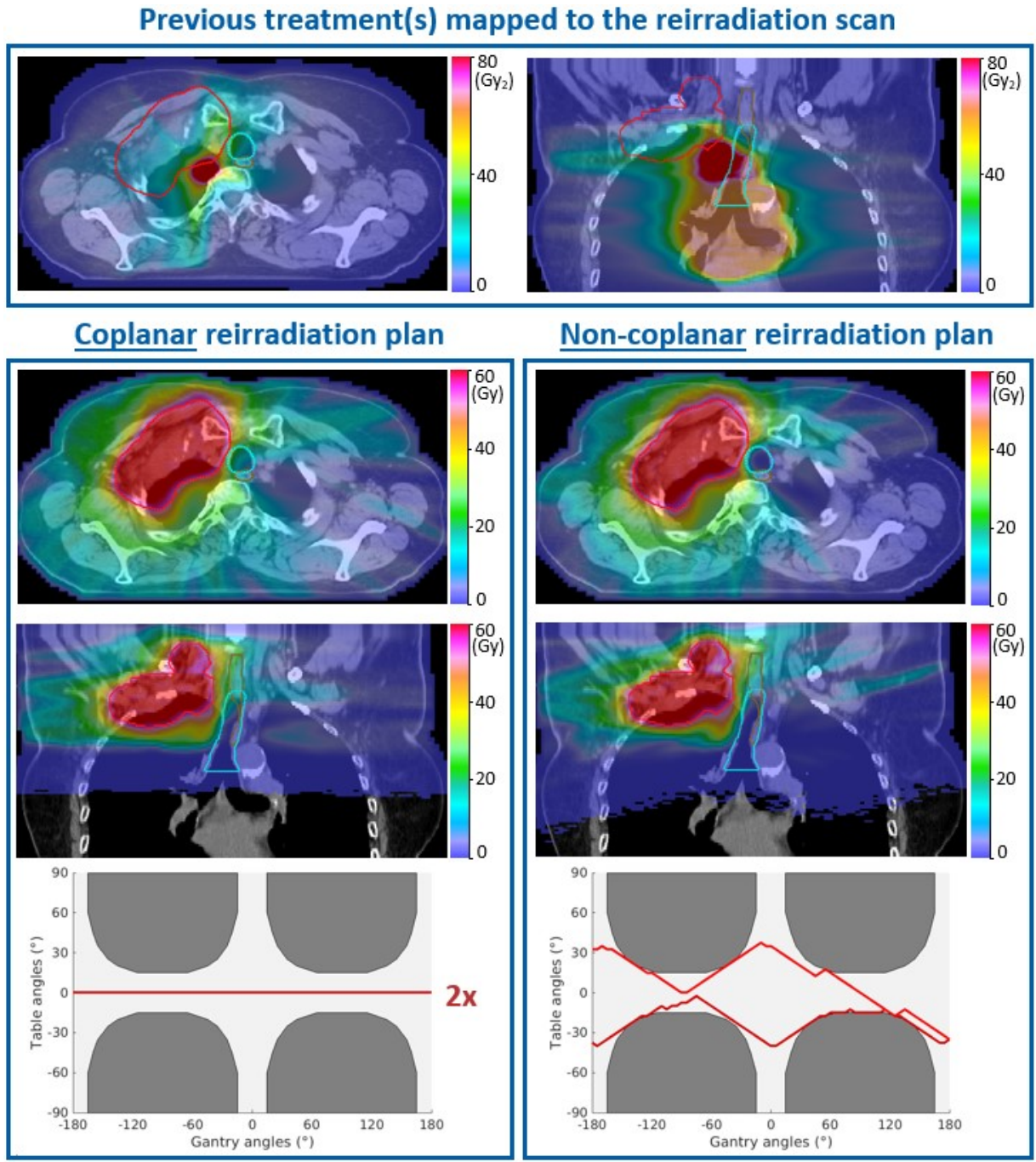

*Figure S8: Comparison of the coplanar and non-coplanar reirradiation plans generated for patient 9. Contours of PTV (red), trachea (light blue) and esophagus (brown) are delineated in both the transversal and coronal planes of the reirradiation scan. The gantry-couch paths for both reirradiation plans are also shown, where dark grey regions indicate beam orientations leading to collision between gantry and couch (and are therefore excluded from the set of candidate beam orientations).*

*Table S9: Cumulative EQD2 metrics for patient 9 achieved both prior to reirradiation and after reirradiation using the coplanar and non-coplanar plans. Percentage differences in dosimetric values between the coplanar and non-coplanar reirradiation plans are reported for each parameter.*

| OAR | EQD2 parameter | Previous treatment(s) | Coplanar reirradiation plan | Non-coplanar reirradiation plan |
|---|---|---|---|---|
| Esophagus | $D_{max}$ (Gy$_2$) | 83.6 | 84.8 | 83.7 (-1.3%) |
|  | $D_{mean}$ (Gy$_2$) | 33.3 | 35.4 | 34.8 (-1.7%) |
| Heart | $D_{max}$ (Gy$_2$) | 66.7 | 66.7 | 66.7 (=) |
|  | $D_{mean}$ (Gy$_2$) | 7.67 | 7.67 | 7.67 (=) |
| Spinal cord | $D_{max}$ (Gy$_2$) | 33.2 | 37.1 | 33.5 (-9.7%) |
|  | $D_{mean}$ (Gy$_2$) | 7.51 | 9.33 | 90.5 (-3.0%) |
| Right brachial plexus[#] | $D_{max}$ (Gy$_2$) | 3.47 | 50.3 | 44.6 (-11.3%) |
|  | $D_{mean}$ (Gy$_2$) | 1.40 | 19.2 | 12.9 (-32.8%) |
| Trachea[#] | $D_{max}$ (Gy$_2$) | 91.1 | 102.7 | 97.4 (-5.2%) |
|  | $D_{mean}$ (Gy$_2$) | 40.2 | 46.0 | 42.5 (-7.6%) |
| Lungs-GTV | $D_{mean}$ (Gy$_2$) | 12.2 | 13.4 | 13.2 (-1.5%) |
|  | $V_{5Gy2}$ (%) | 57.3 | 57.4 | 57.4 (=) |
|  | $V_{20Gy2}$ (%) | 21.3 | 22.7 | 22.5 (-0.9%) |
| PTV | HI | - | 0.14 | 0.15 (+7.1%) |
|  | CI | - | 0.49 | 0.54 (+10.2%) |

[#] OARs which overlap with the PTV

## B.10 Patient 10

The dose distributions along with the corresponding gantry-couch paths obtained for patient 10 are reported in Figure S9 for both the coplanar and non-coplanar reirradiation plans. The dosimetric results are instead detailed in Table S10.

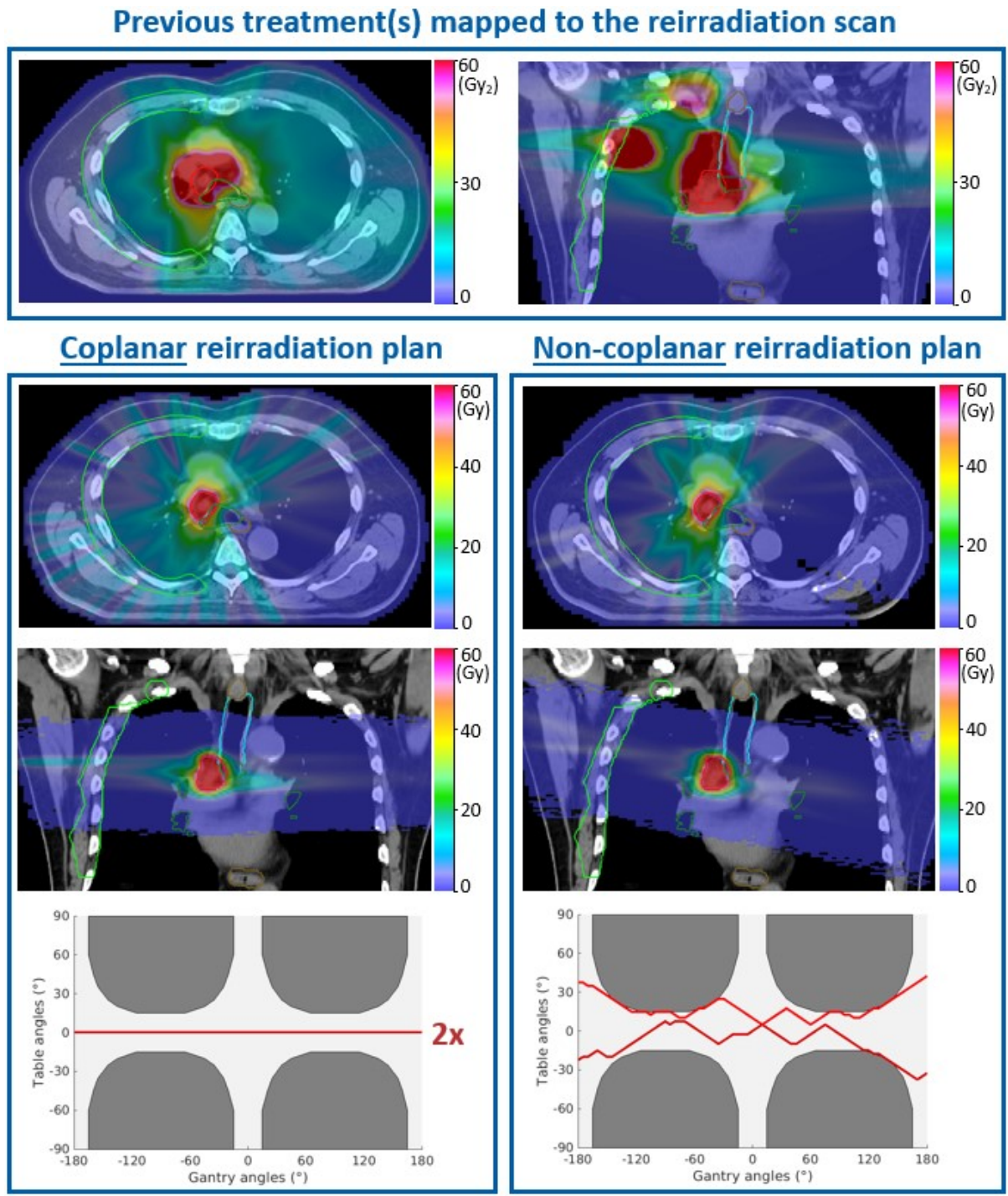

*Figure S9: Comparison of the coplanar and non-coplanar reirradiation plans generated for patient 10. Contours of PTV (red), trachea (light blue), bronchial tree (dark green), thoracic wall (light green) and esophagus (brown) are delineated in both the transversal and coronal planes of the reirradiation scan. The gantry-couch paths for both reirradiation plans are also shown, where dark grey regions indicate beam orientations leading to collision between gantry and couch (and are therefore excluded from the set of candidate beam orientations).*

*Table S10: Cumulative EQD2 metrics for patient 10 achieved both prior to reirradiation and after reirradiation using the coplanar and non-coplanar plans. Percentage differences in dosimetric values between the coplanar and non-coplanar reirradiation plans are reported for each parameter.*

| OAR | EQD2 parameter | Previous treatment(s) | Coplanar reirradiation plan | Non-coplanar reirradiation plan |
|---|---|---|---|---|
| Bronchial tree[#] | $D_{max}$ (Gy$_2$) | 70.4 | 121.2 | 121.2 (=) |
| | $D_{mean}$ (Gy$_2$) | 24.3 | 31.1 | 30.1 (-3.2%) |
| Esophagus | $D_{max}$ (Gy$_2$) | 65.1 | 65.2 | 65.2 (=) |
| | $D_{mean}$ (Gy$_2$) | 7.90 | 7.95 | 7.94 (-0.1%) |
| Heart | $D_{max}$ (Gy$_2$) | 1.30 | 1.31 | 1.31 (=) |
| | $D_{mean}$ (Gy$_2$) | 0.45 | 0.45 | 0.45 (=) |
| Spinal cord | $D_{max}$ (Gy$_2$) | 50.0 | 50.6 | 51.0 (+0.8%) |
| | $D_{mean}$ (Gy$_2$) | 5.46 | 6.02 | 6.09 (+1.2%) |
| Thoracic wall | $D_{max}$ (Gy$_2$) | 173.9 | 173.9 | 173.9 (=) |
| | $D_{mean}$ (Gy$_2$) | 11.2 | 12.2 | 12.1 (-0.8%) |
| Trachea[#] | $D_{max}$ (Gy$_2$) | 71.3 | 115.1 | 111.4 (-3.2%) |
| | $D_{mean}$ (Gy$_2$) | 24.2 | 25.5 | 25.1 (-1.6%) |
| Lungs-GTV | $D_{mean}$ (Gy$_2$) | 8.25 | 9.16 | 9.05 (-1.2%) |
| | $V_{5Gy2}$ (%) | 35.5 | 35.9 | 35.9 (=) |
| | $V_{20Gy2}$ (%) | 11.1 | 13.5 | 12.8 (-5.2%) |
| PTV | HI | - | 0.07 | 0.09 (+28.6%) |
| | CI | - | 0.48 | 0.47 (-2.1%) |

[#] OARs which overlap with the PTV

## B.11 Patient 11

The dose distributions along with the corresponding gantry-couch paths obtained for patient 11 are reported in Figure S10 for both the coplanar and non-coplanar reirradiation plans. The dosimetric results are instead detailed in Table S11.

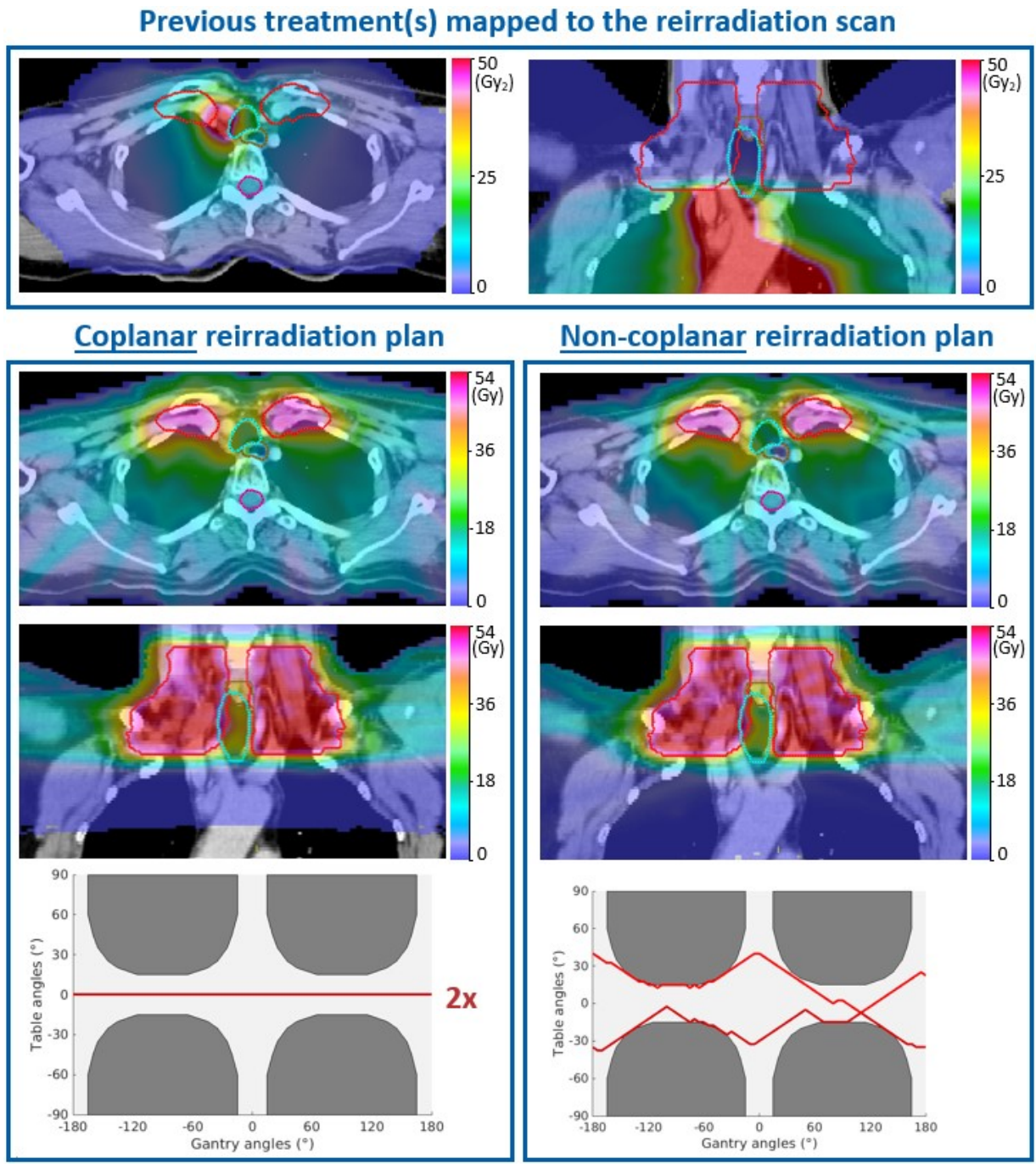

*Figure S10: Comparison of the coplanar and non-coplanar reirradiation plans generated for patient 11. Contours of PTV (red), trachea (light blue), spinal cord (pink) and esophagus (brown) are delineated in both the transversal and coronal planes of the reirradiation scan. The gantry-couch paths for both reirradiation plans are also shown, where dark grey regions indicate beam orientations leading to collision between gantry and couch (and are therefore excluded from the set of candidate beam orientations).*

Table S11: Cumulative EQD2 metrics for patient 11 achieved both prior to reirradiation and after reirradiation using the coplanar and non-coplanar plans. Percentage differences in dosimetric values between the coplanar and non-coplanar reirradiation plans are reported for each parameter.

| OAR | EQD2 parameter | Previous treatment(s) | Coplanar reirradiation plan | Non-coplanar reirradiation plan |
|---|---|---|---|---|
| Esophagus[#] | $D_{max}$ (Gy$_2$) | 69.4 | 69.4 | 69.4 (=) |
| | $D_{mean}$ (Gy$_2$) | 25.0 | 36.7 | 32.7 (-10.9%) |
| Spinal cord | $D_{max}$ (Gy$_2$) | 19.0 | 19.5 | 20.7 (+6.2%) |
| | $D_{mean}$ (Gy$_2$) | 4.35 | 7.98 | 7.44 (-6.8%) |
| Thyroid[#] | $D_{max}$ (Gy$_2$) | 0.41 | 39.4 | 30.9 (-21.6%) |
| | $D_{mean}$ (Gy$_2$) | 0.27 | 12.3 | 7.34 (-40.3%) |
| Trachea[#] | $D_{max}$ (Gy$_2$) | 69.9 | 84.6 | 79.5 (-6.0%) |
| | $D_{mean}$ (Gy$_2$) | 18.8 | 35.4 | 30.8 (-13.0%) |
| Right brachial plexus[#] | $D_{max}$ (Gy$_2$) | 0.64 | 31.9 | 22.5 (-29.5%) |
| | $D_{mean}$ (Gy$_2$) | 0.58 | 14.2 | 7.1 (-50.0%) |
| Left brachial plexus[#] | $D_{max}$ (Gy$_2$) | 0.39 | 27.9 | 21.5 (-1.2%) |
| | $D_{mean}$ (Gy$_2$) | 0.33 | 10.1 | 5.57 (-44.9) |
| PTV | HI | - | 0.24 | 0.29 (+20.8%) |
| | CI | - | 0.75 | 0.74 (-1.3%) |

[#] OARs which overlap with the PTV

## B.12 Patient 12

The dose distributions along with the corresponding gantry-couch paths obtained for patient 12 are reported in Figure S11 for both the coplanar and non-coplanar reirradiation plans. The dosimetric results are instead detailed in Table S12.

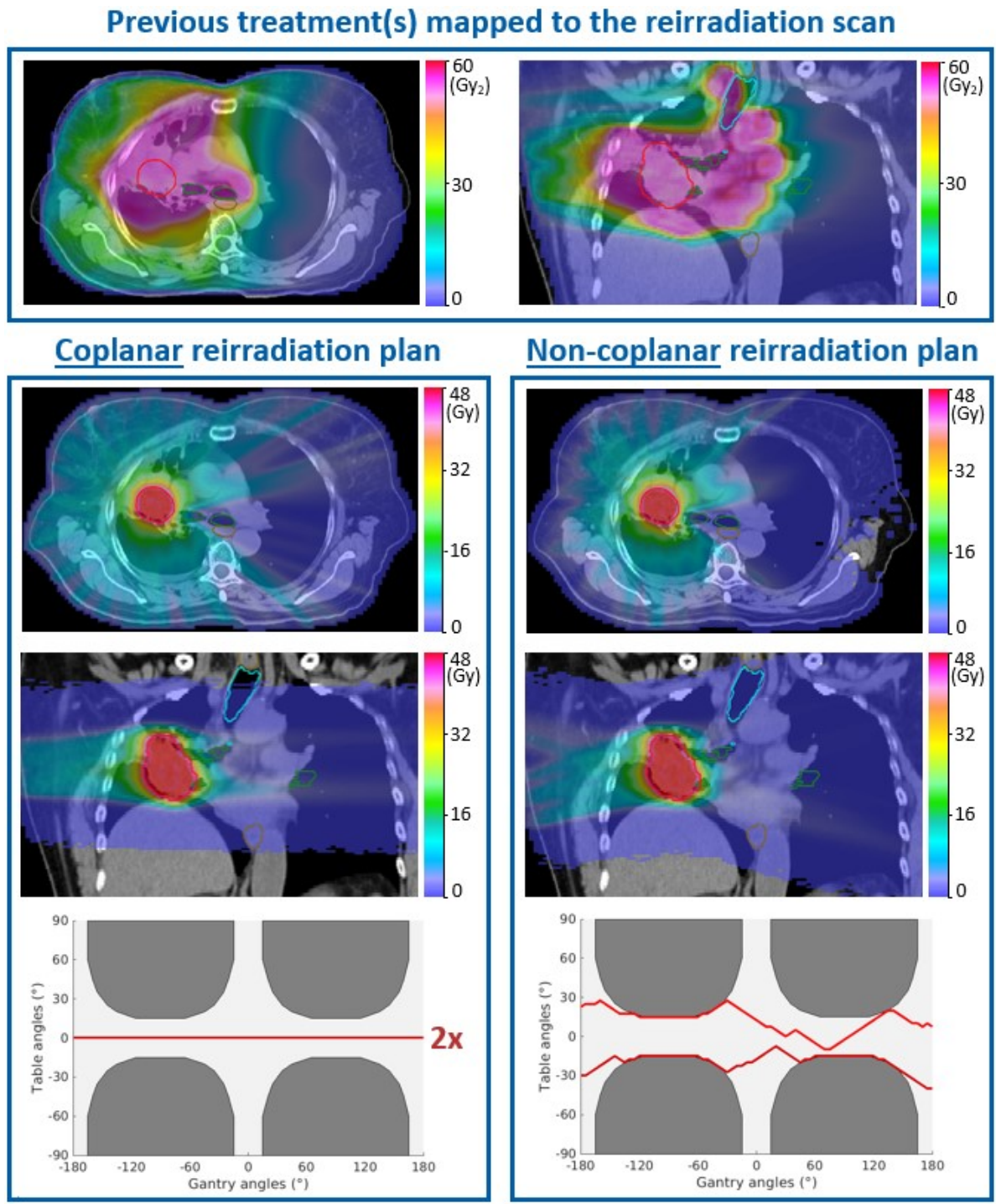

*Figure S11: Comparison of the coplanar and non-coplanar reirradiation plans generated for patient 12. Contours of PTV (red), trachea (light blue), bronchial tree (green) and esophagus (brown) are delineated in both the transversal and coronal planes of the reirradiation scan. The gantry-couch paths for both reirradiation plans are also shown, where dark grey regions indicate beam orientations leading to collision between gantry and couch (and are therefore excluded from the set of candidate beam orientations).*

*Table S12: Cumulative EQD2 metrics for patient 12 achieved both prior to reirradiation and after reirradiation using the coplanar and non-coplanar plans. Percentage differences in dosimetric values between the coplanar and non-coplanar reirradiation plans are reported for each parameter.*

| OAR | EQD2 parameter | Previous treatment(s) | Coplanar reirradiation plan | Non-coplanar reirradiation plan |
|---|---|---|---|---|
| Bronchial tree# | $D_{max}$ (Gy$_2$) | 58.2 | 112.3 | 104.1 (-7.3%) |
|  | $D_{mean}$ (Gy$_2$) | 45.5 | 51.3 | 49.0 (-4.5%) |
| Esophagus | $D_{max}$ (Gy$_2$) | 58.1 | 58.8 | 58.3 (-0.9%) |
|  | $D_{mean}$ (Gy$_2$) | 28.2 | 28.5 | 28.3 (-0.7%) |
| Heart | $D_{max}$ (Gy$_2$) | 57.3 | 64.1 | 63.4 (-1.1%) |
|  | $D_{mean}$ (Gy$_2$) | 9.92 | 10.2 | 10.4 (+2.0%) |
| Spinal cord | $D_{max}$ (Gy$_2$) | 26.5 | 28.3 | 27.4 (-3.2%) |
|  | $D_{mean}$ (Gy$_2$) | 7.88 | 8.31 | 8.36 (+0.6%) |
| Trachea | $D_{max}$ (Gy$_2$) | 58.3 | 59.7 | 58.3 (-2.3%) |
|  | $D_{mean}$ (Gy$_2$) | 33.5 | 33.6 | 33.6 (=) |
| Liver-GTV | $D_{max}$ (Gy$_2$) | 57.9 | 111.1 | 108.5 (-2.3%) |
|  | $D_{mean}$ (Gy$_2$) | 4.41 | 4.54 | 4.53 (-0.2%) |
| Lungs-GTV | $D_{mean}$ (Gy$_2$) | 14.7 | 17.1 | 16.8 (-1.8%) |
|  | $V_{5Gy2}$ (%) | 55.2 | 59.9 | 56.4 (-5.8%) |
|  | $V_{20Gy2}$ (%) | 24.5 | 25.4 | 25.3 (-0.4%) |
| PTV | HI | - | 0.13 | 0.13 (=) |
|  | CI | - | 0.48 | 0.46 (-4.2%) |

# OARs which overlap with the PTV

B.13 Patient 13

The dose distributions along with the corresponding gantry-couch paths obtained for patient 13 are reported in Figure S12 for both the coplanar and non-coplanar reirradiation plans. The dosimetric results are instead detailed in Table S13.

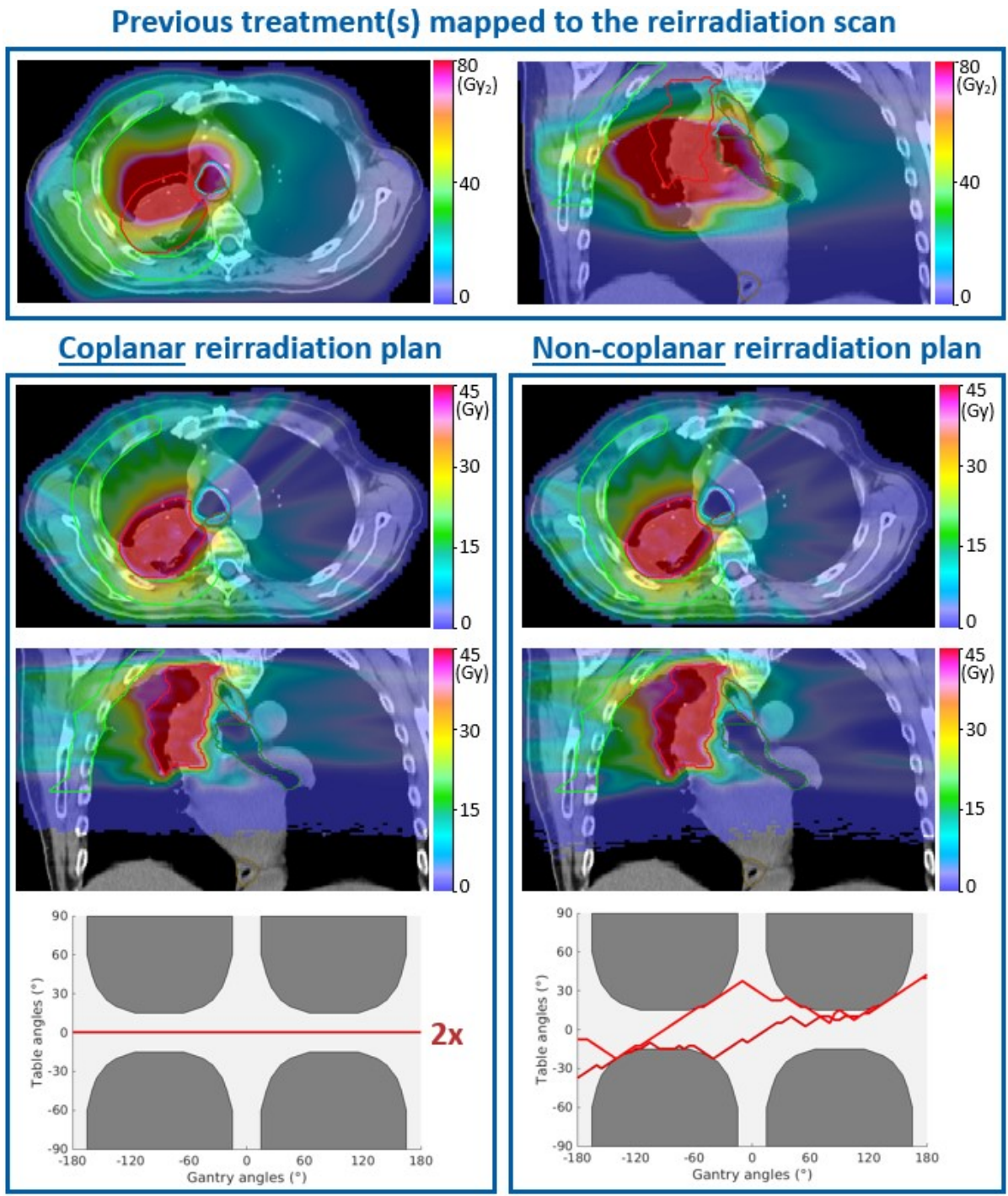

*Figure S12: Comparison of the coplanar and non-coplanar reirradiation plans generated for patient 13. Contours of PTV (red), trachea (light blue), bronchial tree (dark green), thoracic wall (light green) and esophagus (brown) are delineated in both the transversal and coronal planes of the reirradiation scan. The gantry-couch paths for both reirradiation plans are also shown, where dark grey regions indicate beam orientations leading to collision between gantry and couch (and are therefore excluded from the set of candidate beam orientations).*

*Table S13: Cumulative EQD2 metrics for patient 13 achieved both prior to reirradiation and after reirradiation using the coplanar and non-coplanar plans. Percentage differences in dosimetric values between the coplanar and non-coplanar reirradiation plans are reported for each parameter.*

| OAR | EQD2 parameter | Previous treatment(s) | Coplanar reirradiation plan | Non-coplanar reirradiation plan |
|---|---|---|---|---|
| Bronchial tree[#] | $D_{max}$ (Gy$_2$) | 127.4 | 129.2 | 12.90 (-0.2%) |
|  | $D_{mean}$ (Gy$_2$) | 65.7 | 68.5 | 67.9 (-0.9%) |
| Esophagus[#] | $D_{max}$ (Gy$_2$) | 79.2 | 81.4 | 81.1 (-0.4%) |
|  | $D_{mean}$ (Gy$_2$) | 29.6 | 31.5 | 31.6 (+0.3%) |
| Heart | $D_{max}$ (Gy$_2$) | 66.6 | 66.8 | 66.7 (-0.1%) |
|  | $D_{mean}$ (Gy$_2$) | 4.77 | 4.78 | 4.80 (+0.4%) |
| Spinal cord | $D_{max}$ (Gy$_2$) | 24.9 | 30.4 | 28.6 (-5.9%) |
|  | $D_{mean}$ (Gy$_2$) | 4.83 | 6.54 | 6.58 (+0.6%) |
| Thoracic wall[#] | $D_{max}$ (Gy$_2$) | 67.5 | 88.7 | 86.6 (-2.4%) |
|  | $D_{mean}$ (Gy$_2$) | 21.5 | 35.2 | 34.9 (-0.9%) |
| Trachea[#] | $D_{max}$ (Gy$_2$) | 89.4 | 92.6 | 91.0 (-1.7%) |
|  | $D_{mean}$ (Gy$_2$) | 47.3 | 49.8 | 49.0 (-1.6%) |
| Right brachial plexus | $D_{max}$ (Gy$_2$) | 6.89 | 12.8 | 13.4 (+4.7%) |
|  | $D_{mean}$ (Gy$_2$) | 2.17 | 3.29 | 4.10 (+24.6%) |
| Lungs-GTV | $D_{mean}$ (Gy$_2$) | 17.0 | 21.7 | 21.6 (-0.5%) |
|  | $V_{5Gy2}$ (%) | 64.3 | 64.8 | 65.1 (+0.5%) |
|  | $V_{20Gy2}$ (%) | 26.8 | 31.7 | 32.1 (-1.6%) |
| PTV | HI | - | 0.20 | 0.19 (-5.0%) |
|  | CI | - | 0.41 | 0.42 (+2.4%) |

[#] OARs which overlap with the PTV

B.14 Patient 14

The dose distributions along with the corresponding gantry-couch paths obtained for patient 14 are reported in Figure S13 for both the coplanar and non-coplanar reirradiation plans. The dosimetric results are instead detailed in Table S14.

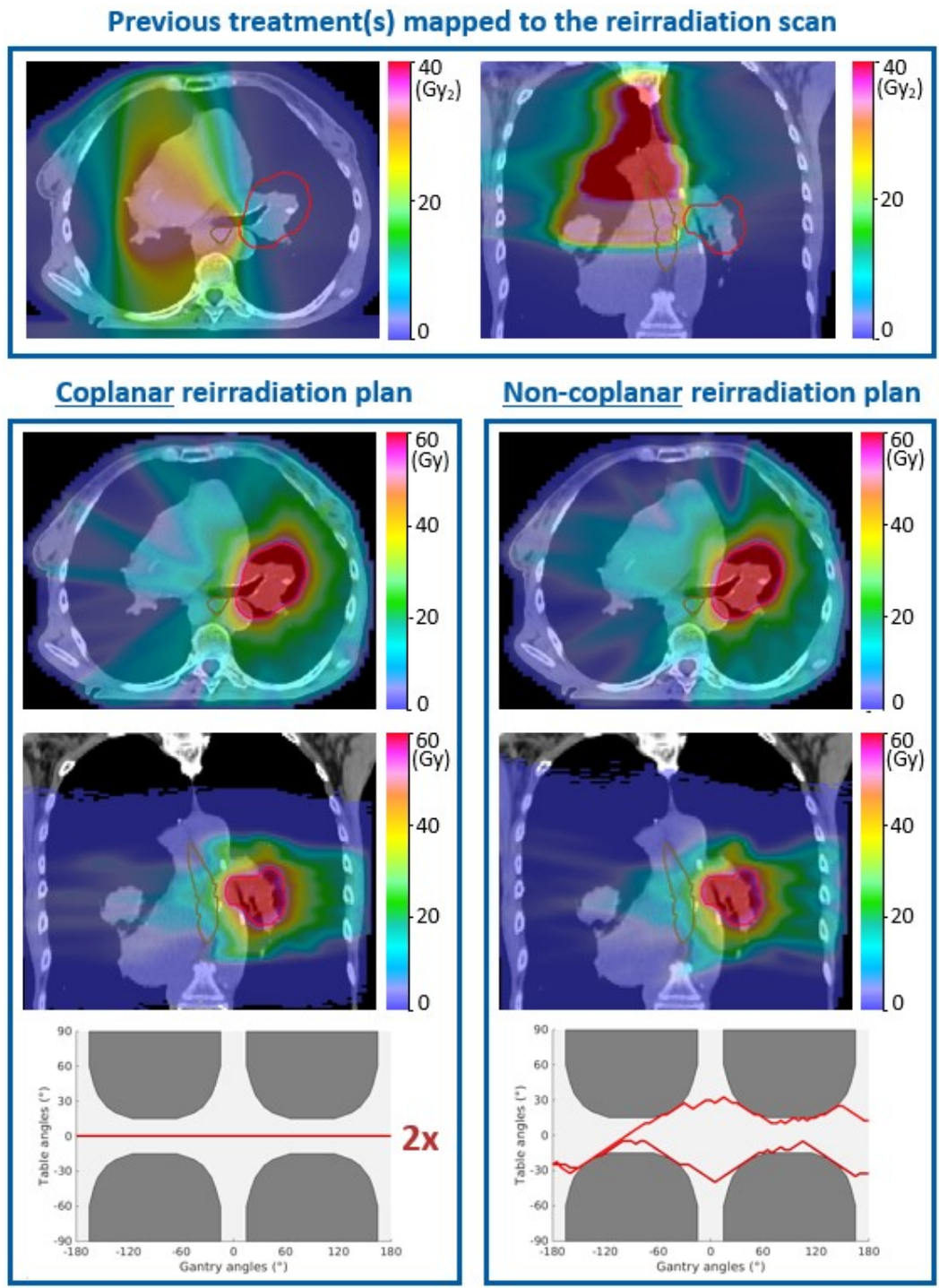

*Figure S13: Comparison of the coplanar and non-coplanar reirradiation plans generated for patient 14. Contours of PTV (red) and esophagus (brown) are delineated in both the transversal and coronal planes of the reirradiation scan. The gantry-couch paths for both reirradiation plans are also shown, where dark grey regions indicate beam orientations leading to collision between gantry and couch (and are therefore excluded from the set of candidate beam orientations).*

Table S14: *Cumulative EQD2 metrics for patient 14 achieved both prior to reirradiation and after reirradiation using the coplanar and non-coplanar plans. Percentage differences in dosimetric values between the coplanar and non-coplanar reirradiation plans are reported for each parameter.*

| OAR | EQD2 parameter | Previous treatment(s) | Coplanar reirradiation plan | Non-coplanar reirradiation plan |
|---|---|---|---|---|
| Esophagus | $D_{max}$ (Gy$_2$) | 63.0 | 63.0 | 63.0 (=) |
|  | $D_{mean}$ (Gy$_2$) | 30.1 | 33.3 | 33.3 (=) |
| Heart[#] | $D_{max}$ (Gy$_2$) | 35.3 | 67.4 | 66.5 (-1.3%) |
|  | $D_{mean}$ (Gy$_2$) | 3.18 | 7.32 | 6.49 (-11.3%) |
| Lungs-GTV | $D_{mean}$ (Gy$_2$) | 7.92 | 13.2 | 12.8 (-3.0%) |
|  | $V_{5Gy2}$ (%) | 45.3 | 64.9 | 63.9 (-1.5%) |
|  | $V_{20Gy2}$ (%) | 12.7 | 23.4 | 21.7 (-7.3%) |
| PTV | HI | - | 0.09 | 0.11 (+22.2%) |
|  | CI | - | 0.48 | 0.48 (=) |

[#] OARs which overlap with the PTV

## B.15 Patient 15

The dose distributions along with the corresponding gantry-couch paths obtained for patient 15 are reported in Figure S14 for both the coplanar and non-coplanar reirradiation plans. The dosimetric results are instead detailed in Table S15.

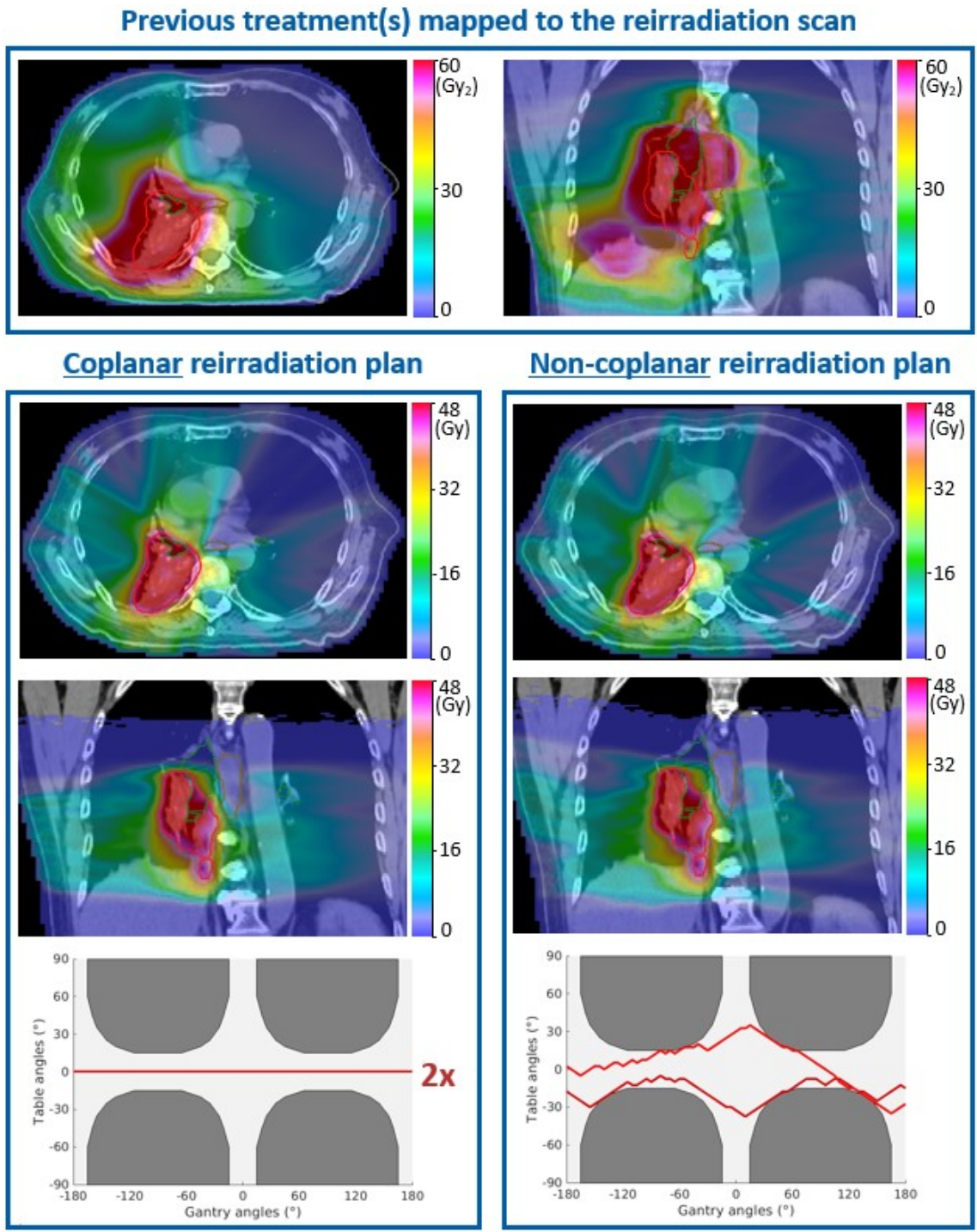

*Figure S14: Comparison of the coplanar and non-coplanar reirradiation plans generated for patient 15. Contours of PTV (red), bronchial tree (green) and esophagus (brown) are delineated in both the transversal and coronal planes of the reirradiation scan. The gantry-couch paths for both reirradiation plans are also shown, where dark grey regions indicate beam orientations leading to collision between gantry and couch (and are therefore excluded from the set of candidate beam orientations).*

*Table S15: Cumulative EQD2 metrics for patient 15 achieved both prior to reirradiation and after reirradiation using the coplanar and non-coplanar plans. Percentage differences in dosimetric values between the coplanar and non-coplanar reirradiation plans are reported for each parameter.*

| OAR | EQD2 parameter | Previous treatment(s) | Coplanar reirradiation plan | Non-coplanar reirradiation plan |
|---|---|---|---|---|
| Bronchial tree[#] | $D_{max}$ (Gy$_2$) | 64.6 | 113.0 | 114.0 (+0.9%) |
| | $D_{mean}$ (Gy$_2$) | 51.6 | 55.1 | 54.8 (-0.5%) |
| Esophagus | $D_{max}$ (Gy$_2$) | 64.4 | 64.8 | 64.7 (-0.2%) |
| | $D_{mean}$ (Gy$_2$) | 21.9 | 23.0 | 23.3 (+1.3%) |
| Heart[#] | $D_{max}$ (Gy$_2$) | 60.6 | 89.4 | 89.1 (-0.3%) |
| | $D_{mean}$ (Gy$_2$) | 5.10 | 9.82 | 9.56 (-2.6%) |
| Spinal cord | $D_{max}$ (Gy$_2$) | 42.5 | 51.6 | 51.8 (+0.4%) |
| | $D_{mean}$ (Gy$_2$) | 12.9 | 14.6 | 14.7 (+0.7%) |
| Trachea | $D_{max}$ (Gy$_2$) | 65.3 | 65.4 | 65.4 (=) |
| | $D_{mean}$ (Gy$_2$) | 27.8 | 27.8 | 27.8 (=) |
| Lungs-GTV | $D_{mean}$ (Gy$_2$) | 12.7 | 15.8 | 15.7 (-0.6%) |
| | $V_{5Gy2}$ (%) | 74.7 | 78.0 | 77.6 (-0.5%) |
| | $V_{20Gy2}$ (%) | 17.8 | 22.9 | 22.3 (-2.6%) |
| PTV | HI | - | 0.18 | 0.17 (-5.6%) |
| | CI | - | 0.43 | 0.43 (=) |

[#] OARs which overlap with the PTV

B.16 All patients

The cumulative EQD2 delivered to the most critical thoracic OARs by both the coplanar and non-colpanar reirradiation plans are summarized for all patients in Figure S15.

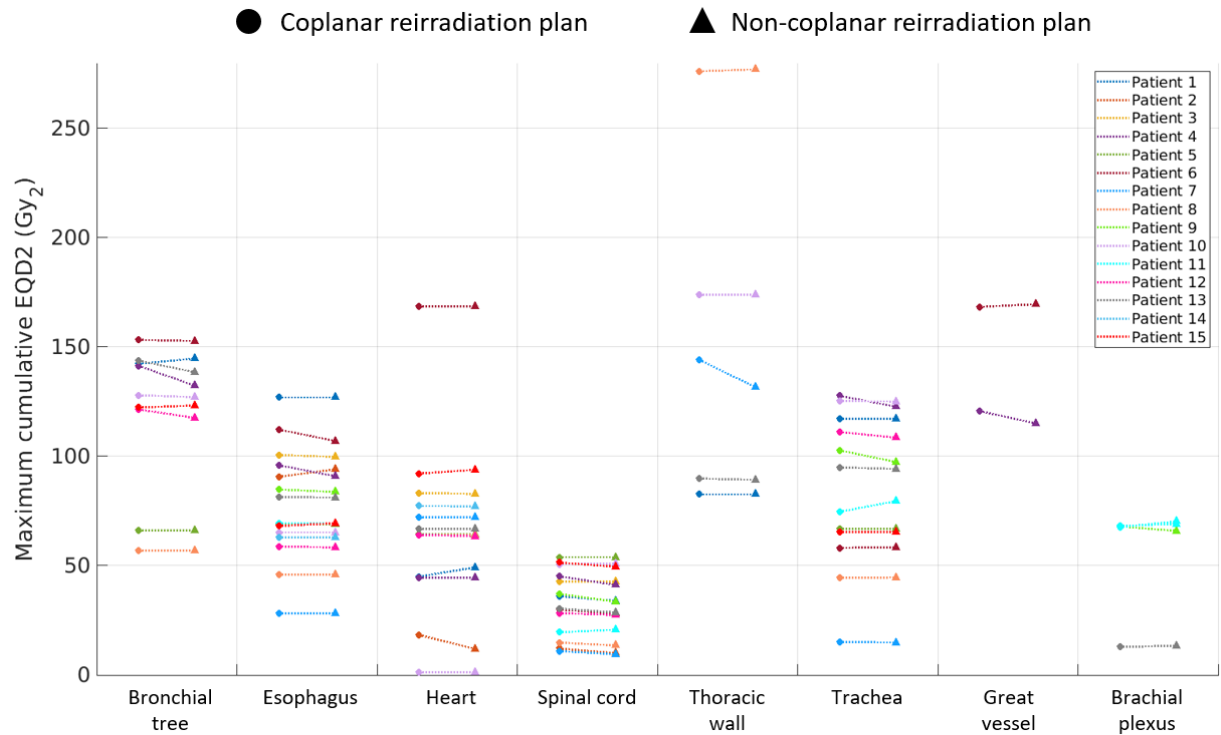

*Figure S15: Comparison of the maximum cumulative EQD2 delivered to the most critical thoracic OARs in all patients using coplanar (round marker) and non-coplanar (triangular marker) reirradiation plans, respectively.*